\documentclass{aastex62}

\usepackage{newtxtext,newtxmath}
\usepackage{float}
\usepackage{color}
\usepackage{xcolor}
\usepackage{booktabs}
\usepackage{multirow}
\usepackage{natbib}
\usepackage{comment}
\usepackage[normalem]{ulem}
\usepackage[T1]{fontenc}
\usepackage{ae,aecompl}
\usepackage{gensymb}

\usepackage{graphicx}	
\usepackage{amsmath}	
\usepackage{amssymb}	

\newcommand{\Ms}{M$_\odot$}

\newcommand{\vir}[1]{``#1''}

\newcommand{\nbody}{\texttt{NBODY7}}

\graphicspath{{./}{figures/}}

\received{XXX}
\revised{YYY}
\accepted{2020}
\submitjournal{ApJ}

\shorttitle{Binary stars in stellar clusters}
\shortauthors{Rastello et al.}

\begin{document}

\title{Effect of binarity in star cluster dynamical mass determination}

\correspondingauthor{Giovanni Carraro}
\email{giovanni.carraro@unipd.it}

\author[0000-0002-5699-5516]{Sara Rastello}
\affiliation{Dipartmento di Fisica, Sapienza, Universit\'a di Roma,\\
 P.le A. Moro 5, I-00185 Roma, Italy}
\affiliation{Dipartimento di Fisica e Astronomia\\
Universit\'a di Padova, Vicolo Osservatorio 3, I-35122\\
Padova, Italy}

\author[0000-0002-0155-9434]{Giovanni Carraro}
\affiliation{Dipartimento di Fisica e Astronomia\\
Universit\'a di Padova, Vicolo Osservatorio 3, I-35122\\
Padova, Italy}

\author[0000-0002-6871-9519]{Roberto Capuzzo-Dolcetta}
\affiliation{Dipartmento di Fisica, Sapienza, Universit\'a di Roma,\\
 P.le A. Moro 5, I-00185 Roma, Italy}

\nocollaboration

\begin{abstract}

 In this paper we explore the effects that the presence of a fraction of binary
stars has in the determination of a star cluster mass via the
virial theorem. To reach this aim in an accurate and consistent way, we run a
set of simulations using the direct summation, high precision, code \nbody.  By
means of this suite of simulations we are able to quantify the overestimate of
open-star-cluster-like models' dynamical masses when making a straight
application of the virial theorem using  available position and radial velocity
measurements. The mass inflation caused by the binary \vir{heating} contribution to
 the measured velocity dispersion depends, of course, on the initial binary fraction, $f_{b0}$  and its following dynamical evolution. For an $f_{b}$(evolved up to 1.5 Gyr) in the range $8\% - 42\%$ the
overestimate of the mass done using experimentally sounding  estimates for the velocity dispersion can be up to a factor $45$. We provide a useful
fitting formula to correct the dynamical mass determination for the presence of
binaries, and underline how neglecting the role of binaries in stellar systems
might lead to erroneous conclusions about their total mass budget. If this
trend remains valid for larger systems like dwarf spheroidal galaxies, which are still far out of reach for high precision dynamical simulations taking account 
their binaries, it would imply an incorrect over-estimation of their
dark matter content, as inferred by means of available velocity dispersion measurements.
\end{abstract}


\section{Introduction} Binary stars play a very important role in the dynamics
of stellar systems, from small open star clusters \citep{hut92} to dwarf
spheroidal galaxies \citep{spencer17}. There are both observational and
numerical indications that binaries observed in stellar systems (from
loose open clusters to very dense globular clusters) cannot be entirely
explained by dynamical formation processes, such as three-body dynamics or
two$-$body tidal capture \citep{aarsethlecar}, but should be, rather,
primordial \citep{hut92,spz01,bt,kouwenbin}. This fraction of tight
primordial binaries may affect dynamical mass estimates most.

The dynamical evolution of a star cluster depends strongly on its binary
population: even a small initial binary fraction can play a fundamental role in
governing cluster dynamics and the whole cluster stellar evolution (see for
instance \citealt{goodman89,mcmillan90,mcmillan91,hut92,mathieu94,
mcmillan94,spz01,goodwin2005,kon, kouwenbin}, and references therein). 
However, the influence of binaries on properties like the bulk motion and
velocity dispersion of stellar systems has not yet fully characterized and
understood.  From an observational point of view it is a challenging task to
catch the entire binary population of a stellar system, because binaries can
have very different periods \citep{geller}, and this induces a clear bias
in the determination of the binary fraction in a cluster if only one or a
few epochs are observed, as it is customary. 
Usually, binaries can be detected by one of the following ways: by spectroscopy
(from radial$-$velocity variations), or photometry  (from the abnormal location
on Hertzsprung${-}$Russell diagram). Other types of binaries are the so
called \vir{visual} binaries (stars too close on the sky to be explained by
chance projection, \cite{goodwin})  and the astrometric binaries (visual
binaries which we see orbiting) .  Clearly, all of these methods are biased;
the first method is biased to close similar$-$mass companions while the second
and the third method are biased towards similar-luminosity or mass (low$-$mass)
companions \citep{goodwin2005,goodwin}.

In globular clusters, the fraction of binaries is regulated by two competing
effects: the formation of bound systems due to dynamical segregation, and their
destruction due to strong dynamical interactions in the cluster cores
\citep{bregman1,bregman,bregman2}.

In looser systems, like open clusters or dwarf galaxies, strong dynamical
interaction are rare, and primordial binaries may survive longer \citep{spz01}.
The fraction of primordial binaries in open clusters is estimated to range
between $30\%$ and $60$\% \citep{spz01,sana2008, sana2009,sana2011,sana2013}
approaching $100\%$\% in some particular cases
\citep{fan96,sana2008,fuente2008}.

In dwarf galaxies, in particular, the presence of binary stars has been
considered as a potential explanation for the difference in velocity dispersion
with respect to globular clusters of comparable mass. Actually, the velocity
dispersion in dwarf galaxies is, typically, larger, and also the mass to light
ratio, which would imply a very large content of dark matter
\citep{spencer17}.

The possible link between dark matter content and binary fraction is
particularly intriguing.  Globular clusters are dark matter free, and host
small fractions of binaries. On the other side, dwarf spheroidal galaxies
\citep{spencer17} seem to be dark matter dominated and, at the same time, they
host many binaries.  The bridge between dark matter content and binary content
could be the velocity dispersion, via the virial theorem. Previous studies
showed that when binary stars are properly taken into account the velocity
dispersion estimation and, in turn, the evaluation of the virial mass of dwarf galaxies tend to decrease \citep{spencer17,spencer18}.

Therefore, the knowledge of the binary content of a stellar system allows
clearing the bias they induce: the better they are taken into account, the
smaller the velocity dispersion results to be \citep{kouwenbin,spencer17}.
This issue is particularly crucial for low density systems, like Bootes I
\citep{munoz,kosopov}, or Segue 1 \citep{geha,simon, martinez}.

Direct measurements of the 1D velocity dispersion ($\sigma_{los}$) of a star
cluster can be done in three different ways: ($i$) estimating the width of spectral lines
from observations spanning a significant part of the cluster \citep{moll2009};
($ii$) measuring the radial velocities of individual stars \citep{apai2007}
and, ($iii$) proper motions \citep{chenl2007,toff2014}, possible only for
nearby clusters. Due to the nature of the observations, the velocity dispersion
obtained using techniques ($i$) and ($ii$) might be significantly affected by
the presence of binaries \citep{kouwenbin}, while  proper motion
measurements do not lead to a mass overestimation, even when the binary
fraction is high. For instance, single-epoch velocity dispersion is larger
than multi-epoch velocity dispersion, and the wider the time coverage, the
smaller is the resulting velocity dispersion. This means that binary stars can
\vir{inflate} the velocity dispersion of stellar systems.  In fact, in a
cluster consisting only of single stars, the velocity dispersion strictly
correlates with the motion of each \vir{particle} in the cluster potential. On
the other hand, in a cluster populated also by binary stars we cannot easily
pick the motion of the binary center of mass from that of the individual binary
components which gives two additional degrees of freedom (like roto-vibrations
in a diatomic molecule),  the rotational one being the dominant. This may
induce, thus, an overestimation of the dynamical cluster mass
\citep{fleck,apai2007,gieles2010} as computed using 
the virial theorem.

Given all the above, we note that, interestingly,  the mean stellar density
and binary fraction of dwarf spheroidal galaxies is comparable to that of open
clusters in the Milky Way disk \citep{Kharchenko,McConnachie,spencer18}. It is
therefore tempting to start testing systematically the binary effect on open
clusters first, given the obvious numerical advantage, to look for similarities
and/or differences with observations.

Open clusters are in fact small enough to allow performing multiple simulations
of their dynamics at a level sufficient to give a good statistical coverage of
their properties, yet they are large enough and old enough that both stellar
evolution and stellar dynamics have had time to play significant roles in
determining their present structure. \citep{spz01}.

Moreover, open clusters contain fractions of binaries larger than globulars
\citep{fuente2008,carrarooc}, and are widely accepted to be devoid of dark
matter halos.  The small number of stars belonging to open clusters allows a
tight comparison with numerical models.

Generally, high precision dynamical models studied so far, exclude binaries for
simple practical reasons \citep{mikkola98,spz01,trenti}: ($i$)
binaries slow down calculations dramatically and induce huge numerical errors;
($ii$) their internal evolution is much more complicated than the evolution of
single stars; ($iii$) a good treatment of binaries would require accurate
dynamical regularization tools.

However from a theoretical/numerical side, given the relatively small number of
open clusters' member stars it is easy to explore the role of binaries in such
systems.

In this paper we address the role of binaries in open cluster-like stellar
systems in influencing the cluster velocity dispersion and thus the
determination of the cluster \vir{dynamical} mass.  Our work bases on high
precision, direct summation, $N$-body simulations. With the aim to span a wide
range of initial conditions, we model open clusters at varying the initial
fraction of primordial binaries and the cluster initial \textit{virial} state
(by mean of varying the initial virial ratio $Q=2T/|\Omega|$ where $T$ and
$\Omega$ are the total kinetic and potential energy, respectively; $Q=1$
corresponds to virial equilibrium). Although this is the most
straightforward definition of a virial ratio, coming directly from the
expression of the 2nd time derivative of the polar moment of inertia of a
system of $N$ gravitating objects, we note that some papers refer to
\textit{virial} ratio as the $Q=T/|\Omega|$ ratio, that gives 1/2 for a
virialised system. To estimate the velocity dispersion of the cluster ,
and hence the system kinetic energy,  we use three different
methods, accounting in different ways for the presence of binary stars.

The paper is organized as follows: in Sect. 2 we describe the models and the
numerical methods we used; in Sect. 3 we present and discuss our results.
Summary and conclusions are drawn in Sect. 4.

\section{Method and Models}
\label{mm}
\subsection{$N$-body models}
\label{sim}

In order to study the effect of binaries on the estimation of the dynamical
mass of open clusters we made use of high precision direct $N$-body simulations
performed with the code \texttt{NBODY7} developed by \citet{asnitadori12}.
\texttt{NBODY7} is a direct summation $N$-body code that integrates in a
reliable way the motion of stars in not too abundant stellar systems and which
implements sophisticated and efficient recipes to deal with strong
gravitational encounters, taking also into account stellar evolution. The high
precision treatment of binary stars is allowed in \texttt{NBODY7} thanks to the
KS regularization tool \citep{ks} and the Algorithmic Regularization Chain
(ARC) \citep{archain}.\\
We defined six simulation groups (A, B, C, D, E, and F) representing various
open clusters at varying the population of primordial binaries and the cluster
virial state. All the clusters contain initially $N_0\,=\,1000$ stars. In all
cases, the radial distribution was set according to a Plummer density profile
\citep{Plum}. The initial total mass of each cluster is, in dependence on the
random seed for the sampling, in the interval   $600 - 700$  \Ms. The
cluster core radius is $r_{c} = 1$ pc, and for each system we adopt solar
metallicity (Z$_{\odot}$). We assume a \citet{kroupa01} initial mass function
with masses in the range $0.01$ \Ms $\leq m \leq$ $100$ \Ms.  The clusters
are supposed to be isolated and considered as proxies of open clusters of the
Milky Way.

For each model (A, B, C, D, E, and F) we vary the fraction of
primordial binaries, $f_{b0}$, and the initial virial ratio, $Q_0$ (see Table
\ref{tab:1}).

The first three groups of simulations (A, B and C) represent clusters in an
initial virial equilibrium ($Q_0\,=\,1$)  while the other groups (D, E and
F) refer to \textit{sub-virial} star clusters, assuming $Q_0=1/2$.

We consider for each cluster model a primordial population of binaries in a
fraction which varies in the $5\%-30\%$ range (Table \ref{tab:1}).  The initial fraction of binaries, $f_{b0}$ is defined as the ratio of the initial
number of pairs of stars, $N_{b0}$, to the initial total number of cluster stars, $N_0$, so that $f_{b0}=N_{b0}/N_0$, and $N_0=N_{s0}+2N_{b0}$ is the total number
of stars, provided $N_s$ as number of \textit{single} stars in the system.

The mass ratio distribution of the primordial binary population is modelled
according to the law $f(m_{A}/m_{B})\,\propto (m_{A}/m_{B})^{+0.4}$
(where $m_{A}\geq m_{B}$ are the masses of the two stars in the
binary) \citep{kouvpairing}, while periods are distributed according a
logarithmic distribution \citep{kroupa95} and, for eccentricities ($e$), we
assumed a ``thermal'' distribution $g(e)= 2e$ \citep{jeans19}.  For each model,
we run $10$ $N$-body simulations. Each simulation is a different realization of
each cluster model for which we change the random \vir{seed} when creating the initial
conditions.

In particular, the initial conditions drawn this way are obtained updating the
procedure followed in \cite{ASCD15He}.  We evolved all the models up to $\sim$
$1.5$ Gyr. All the simulations were performed with the multi GPU workstation
ASTROC16A hosted at Sapienza, University of Rome.

\begin{table}[!h]
\centering
\begin{tabular}{@{}ccc@{}}
\toprule \multicolumn{1}{l}{{Model}} &\multicolumn{1}{l}{$f_{b0}$ (\%)} & \multicolumn{1}{l}{$Q_0$} \\
\midrule
A                                  & 5 & 1                                             \\
B                                  & 15 & 1                                             \\
C                                  & 30 & 1                                             \\ \midrule
D                                  & 5 & 0.5                                           \\
E                                  & 15 & 0.5                                           \\
F                                  & 30 & 0.5                                           \\ \bottomrule
\end{tabular}
\caption{Summary of the initial conditions adopted in our models.  The
columns from left to right represent: 1) the label of the model, 2) the  percentage fraction of primordial binaries ($f_{b0}$), 3) the initial virial
ratio ($Q_0$) of the clusters.}
\label{tab:1}
\end{table}

\subsection{Strategy}
\label{stra}

It is intuitive that binary stars play a relevant role when estimating the mass
of a star cluster by mean of virial considerations. This is because their presence
can affect significantly the \vir{observed} velocity dispersion, giving an
extra contribution over the pure kinetic (translational) one that would
provide the correct evaluation of kinetic energy at the numerator of the virial
ratio $Q$. Of course, any overestimate of the kinetic energy, at a given $Q$
value, leads to a corresponding overestimate of the mass of the system.\\ From
an observational point of view, it is real challenging the disentangling the
binary population, especially when binaries have long periods and small
amplitudes, although long period binaries are the ones that less affect the velocity dispersion measurements because of their lower velocities around the pair barycenter. Typically, multi-epoch and high precision radial velocity measurements are required.  So, to evaluate the binary role it is much more
feasible using a \vir{direct} and controlled approach, that means to build up a
set of $N$-body realisations of a cluster whose binary content is pre-defined,
to get numerical outputs which allow checking how the velocity dispersion
evaluations can be biased. Therefore,
we estimated the cluster 3D velocity dispersion in all of our models
of Table \ref{tab:1} by means of four different methods, described as
follows:

\begin{itemize}

\item Method $1$: the \textit{total} velocity dispersion (hereafter denoted
with $\sigma_{tot}$) is estimated accounting for all the stars of the cluster
as if \vir{they were all single stars}, i.e. independently of possible
binarity. In practice, given $N$ velocity vectors, $\mathbf{v}_i$
($i=1,2,...,N$), we scaled them to the proper rest frame to evaluate the
\textit{total} velocity dispersion

\begin{equation}
\label{sigtot}
\sigma_{tot}=\sqrt{\frac{1}{N}\sum_{i=1}^N\textit{v}_i^2},
\end{equation}

where $\textit{v}_i$ is the absolute value of $\mathbf{v}_i$.

\item Method $2$: here we
make a distinction between the $N_s$ single stars and the $N_b$ binaries, in that, in the velocity dispersion calculation, we consider for every $j$th
($j=1,2,...,N_b$)
binary composed by the two masses $m_{A,j}$ and $m_{B,j}$, only its center of mass velocity,
\begin{equation}
\mathbf{v}_{cm,j}=\dfrac{m_{A,j}\mathbf{v}_{A,j}+m_{B,j}\mathbf{v}_{B,j}}{M_j},
\end{equation}
where $M_j=m_{A,j}+m_{B,j}$ is the binary mass, to evaluate the cluster velocity dispersion as

\begin{equation}
\label{sigcm}
\sigma_{cm}=\sqrt{\frac{1}{N_s+N_b}\left(\sum\limits_{i=1}^{N_s}\textit{v}_i^2+\sum\limits_{j=1}^{N_b}\textit{v}_{cm,j}^2\right)}.
\end{equation}

\item Method $3$: here we keep a distinction between single and binary stars but, in this case, for every binary we consider a luminosity averaged velocity

$$  \mathbf{v}_{lum,j} = \frac{L_{A,j}\mathbf{v}_{A,j} +L_{B,j}\mathbf{v}_{B,j}}{L_j},$$
where $L_j=L_{A,j}+L_{B,j}$ is the binary total bolometric luminosity, so to have a dispersion, $\sigma_{lum}$, defined as

\begin{equation}
\label{siglum}
\sigma_{lum}=\sqrt{\frac{1}{N_s+N_b}\left(\sum\limits_{i=1}^{N_s}\textit{v}_i^2+\sum\limits_{j=1}^{N_b}\textit{v}_{lum,j}^2\right)}.
\end{equation}

\item Method $4$: to have another term of comparison with observations,
we also derived the velocity dispersion over the set of single star only, thus excluding binary systems. The velocity dispersion, referred to as  $\sigma_{sing}$, is so

\begin{equation}
\label{sigsing}
 \sigma_{sing}=\sqrt{\frac{1}{N_s}\sum_{i=1}^{N_s}\textit{v}_i^2}.
\end{equation}

\end{itemize}~

Method $1$ is the simplest possible estimate of the velocity dispersion  but it provides a value of the kinetic energy content which exceeds the actual ($3$ degrees of freedom per particle) contrasting the global potential, because, in a quantity dependent on the binary fraction, it accounts also for the binary inner degrees of freedom which should not be considered in a virial mass determination.  Actually, cluster
observations suffer from the following issues: $i$) for many reasons, we are
able to identify only a fraction of cluster stars, usually the most luminous
stars; $ii$) binaries are difficult to detect and to distinguish respect to
single stars. This is particularly true when dealing with binaries composed by
two stars of significantly different luminosity. In this case it is really hard
to derive the individual velocity components of the two binary stars with high
precision, to pick the center of mass velocity. Thus the kinematic study
requires many data of radial velocity measurements which span a wide range of
time. Consequently, Method $3$ is the closest to what happens when dealing with observations.

As said in Sect. \ref{sim}, the evolution of our open cluster models is
followed up to $\sim 1.5$ Gyr, time at which we estimate the velocity
dispersion according to the methods described above. In order to improve
statistical significance we make averages over a time range $\pm$ $50 $
Myr around $1.5$ Gyr.

\begin{figure}[!h]
\centering
\includegraphics[width=0.5\textwidth]{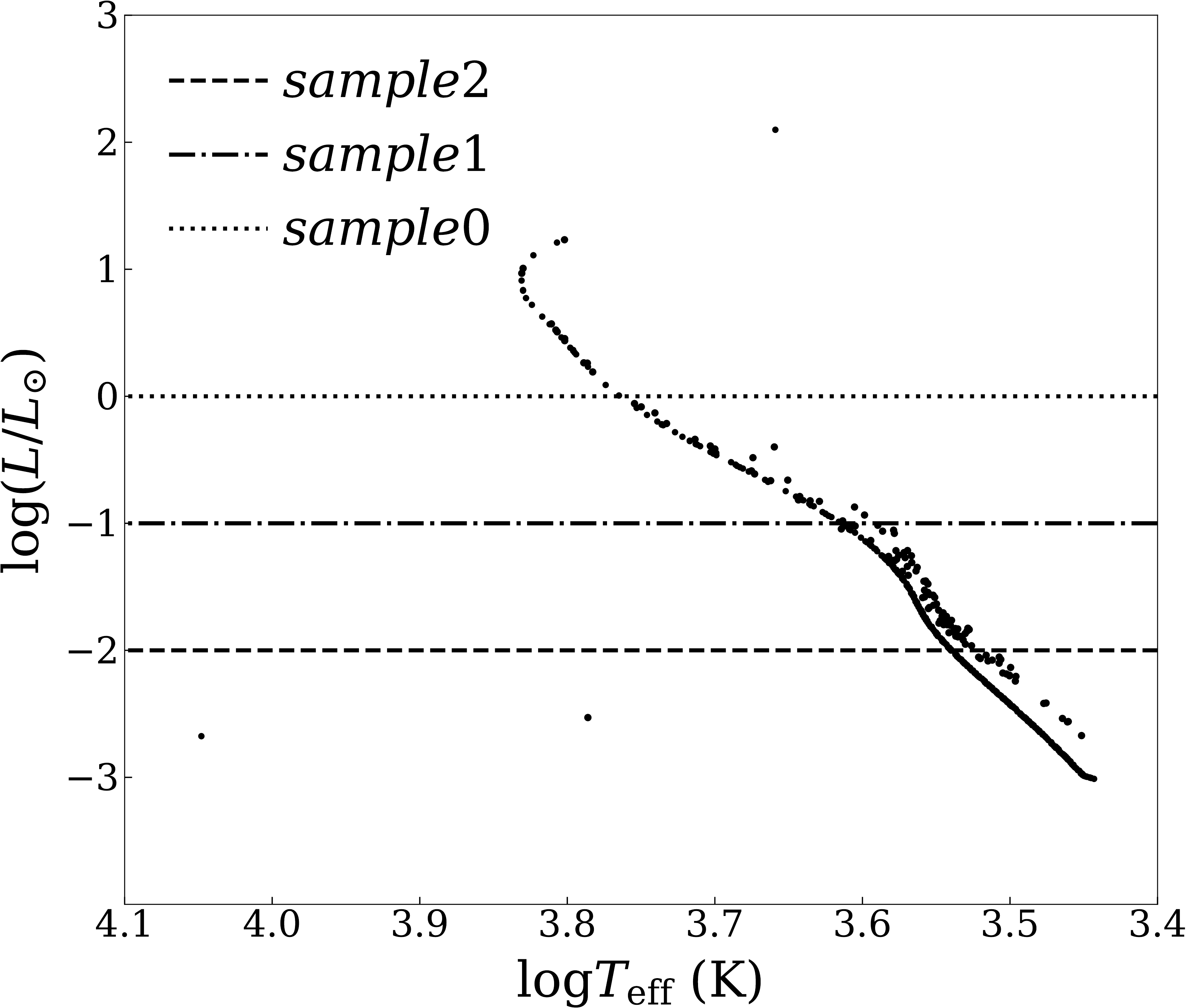}
\caption{Example
of the HR diagram for a simulation of model B at 1.5 Gyr.
The three lines refer to the different luminosity cuts: dashed line for \textit{sample2},
dotted dashed line for \textit{sample1} and dotted line for \textit{sample0}, respectively. The spread
in the Main Sequence track is due to the binaries.}
\label{fig:hrD}
\end{figure}

To give a reliable comparison with observations, we extract three different
samples of stars from each model cluster basing on their luminosity.  We
consider the ($\log$ $T_{\rm{eff}}$, $\log$ $L$) HR diagram as shown in Fig. \ref{fig:hrD}
and select three luminosity biased samples of stars according to the following
thresholds  (named according to the luminosity cut):
 $1$) $\log(L/L_\odot)\, \geq \,-2 $, hereafter addressed as
\textit{sample2}; $2$) $\log(L/L_\odot)\, \geq \, -1$, hereafter addressed as
\textit{sample1} and $3$) $\log(L/L_\odot)\, \geq \, 0$, hereafter addressed as
\textit{sample0}.

Thus, for each model we estimate the velocity dispersion for each selected
sample of luminosity.  The error of the velocity dispersion is evaluated
according to the standard deviation measures.

\section{Results and Discussion}
\label{rr}
\subsection{Cluster Dynamical Evolution}
\label{ev}

Many physical processes influence cluster evolution, among which the most
important are stellar evolution, the Galactic tidal field, and the fraction of
binary and multiple stars. The abundant mass$-$loss from individual star during
their early evolution is of greatest importance and, carrying away much of the
cluster binding energy, it may result in the disruption of the entire cluster
\citep{spz01}.  If the cluster survives this early phase, stellar evolutionary
time$-$scales become longer than the time$-$scales for dynamical evolution,
hence two-body relaxation and tidal effects thus become dominant. Moreover, the
presence of a population of primordial binaries is crucial to both stellar and
dynamical evolution of a cluster \citep{hut92}. Actually, the mass transfer
between binary components allows new stellar evolutionary states to arise and,
in addition, the presence of binaries may enhance the rate and type of stellar
collisions, making possible the temporary capture of single stars and other
binaries in three$-$body encounters \citep{heggie75,spz98}. In addition to the
mass loss due to stellar wind, clusters also lose mass in form of escaping
stars. The fraction of escapers is enhanced if clusters are considered embedded
in the external tidal field of the host galaxy (the escaper rate is estimated
to be of the order of about 10\% per relaxation time \citep{spi87}). The
external tidal field induces truncation of the cluster sizes and lowers the
escape speed, significantly enhancing the mass$-$loss rate \citep{vesperini10}.
This makes the cluster dissolution time significantly shorter than in the
isolated case. We stress here that in our models the clusters are all isolated
systems. Although this assumption is surely questionable for real open
clusters, which are embedded in an external potential and subjected to galactic
differential rotation, it constitutes a needed initial step in this type of
investigations.  Isolated systems undergo mass$-$loss through escapers as due to the
 combined effects of close and distant encounters \citep{heggiebook}.

A summary of the configuration of the clusters is given in
Table \ref{table:config}, where we report the clusters properties after $1.5$ Gyr from the beginning of the simulations. The $\pm$ $50 $ Myr indicates that the results are averaged over a time of $100$ Myr around $1.5$ Gyr.

\begin{table}[!h] 
\centering 
\begin{tabular}{@{}cccc@{}} 
\toprule {Model} &{$\langle N \rangle$} & {$\langle M_{cl} \rangle $ (M$_{\odot}$)} & \textbf{$\langle f_b\rangle $ (\%)} \\ \midrule &                                    &
&                                            \\ A              & $730$
& 290.20                                & 5.6                               \\
&                                    &
&                                            \\ B              & $723$
& 279.40                                & 17.1                              \\
&                                    &
&                                            \\ C              & $706$
& 275.60                                & 35.1                              \\
\midrule &                                    &
&                                            \\ D              & $637$
& 251.87                                & 6.3                               \\
&                                    &
&                                            \\ E              & $614$
& 240.98                                & 18.9                              \\
&                                    &
&                                            \\ F              & $601$
& 240.19                                & 37.9                              \\
\bottomrule 
\end{tabular} 
\caption{Some parameters of the model clusters at t$\,=\,1.5$ Gyr.  From left to right the various columns give: 1)
the model identification label (see Table \ref{tab:1}); 2) the averaged number of bound stars
($\langle N \rangle$); 3) the mean cluster mass ($\langle M_{cl} \rangle
$; 4) the mean percentage of binaries bound to the cluster ($\langle f_b \rangle =\langle N_b/N\rangle$). The reported  values are averaged over all the simulations performed in each set.}
\label{table:config} 
\end{table}~

We notice that in models corresponding to an initial virial equilibrium (A, B
and C) the number of retained stars (and thus the total bound mass $\langle
M_{cl} \rangle$) is larger ($\sim 70$\%) with respect to models on a
initial sub-virial state (D, E and F) after $1.5$ Gyr. Additionally, initially
virialised models show also a somewhat larger fraction of retained binaries:
5\%, 17\% and 35\% which corresponds to $\sim 41$, $\sim 123$ and $\sim 250$
binaries for model A, B, and C, respectively.  In sub-virial models, the effect
of encounters is enhanced and leads to destruction of a large number of
binaries, which are, after $1.5$ Gyr, respectively $\sim 40$, $\sim 117$ and
$\sim 226$ for model D, E and F. The opposite trend in the fraction column of
Table \ref{table:config} is due to that in the sub-virial models the enhanced
ejection of single stars covers the enhanced binary disruption.

A similar results is discussed also in \citet{sana2013}. 
We remind that initially we set $N_{b0}\,=\,50$ for models A
and D, $N_{b0}\,=\,150$ for models B and E and $N_{b0}\,=\,300$ for models C
and F.

Actually, primordial binaries may be disrupted or may exchange components with
other stars, and, in addition, because of the gravitational interactions within
the stellar systems, new binaries may form \citep{hut92}.  However, in open-cluster-like systems strong gravitational
encounters are rare and the disruption of binary systems is generally less
pronounced (apart from wide binaries which are likely to
become unbound) with respect to globular clusters
\citep{Terlevich}. Moreover, binaries are generally heavier than single
stars and tend to segregate toward the central region of the hosting star cluster, where the escape velocity is higher and, so, the probability to escape from the clusters is lower.

\begin{figure}[!h]
\centering
\includegraphics[width=0.45\textwidth]{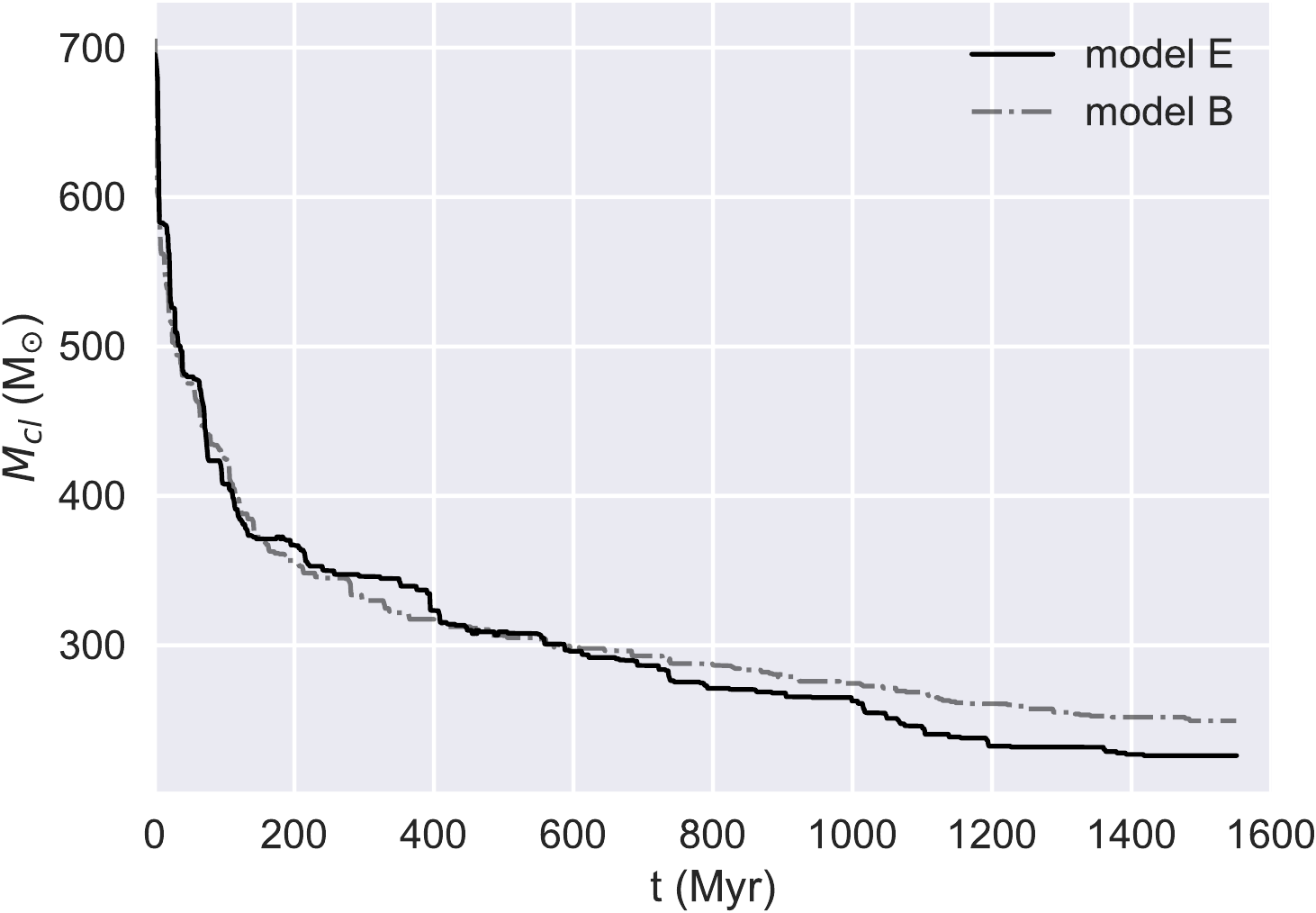}
\includegraphics[width=0.45\textwidth]{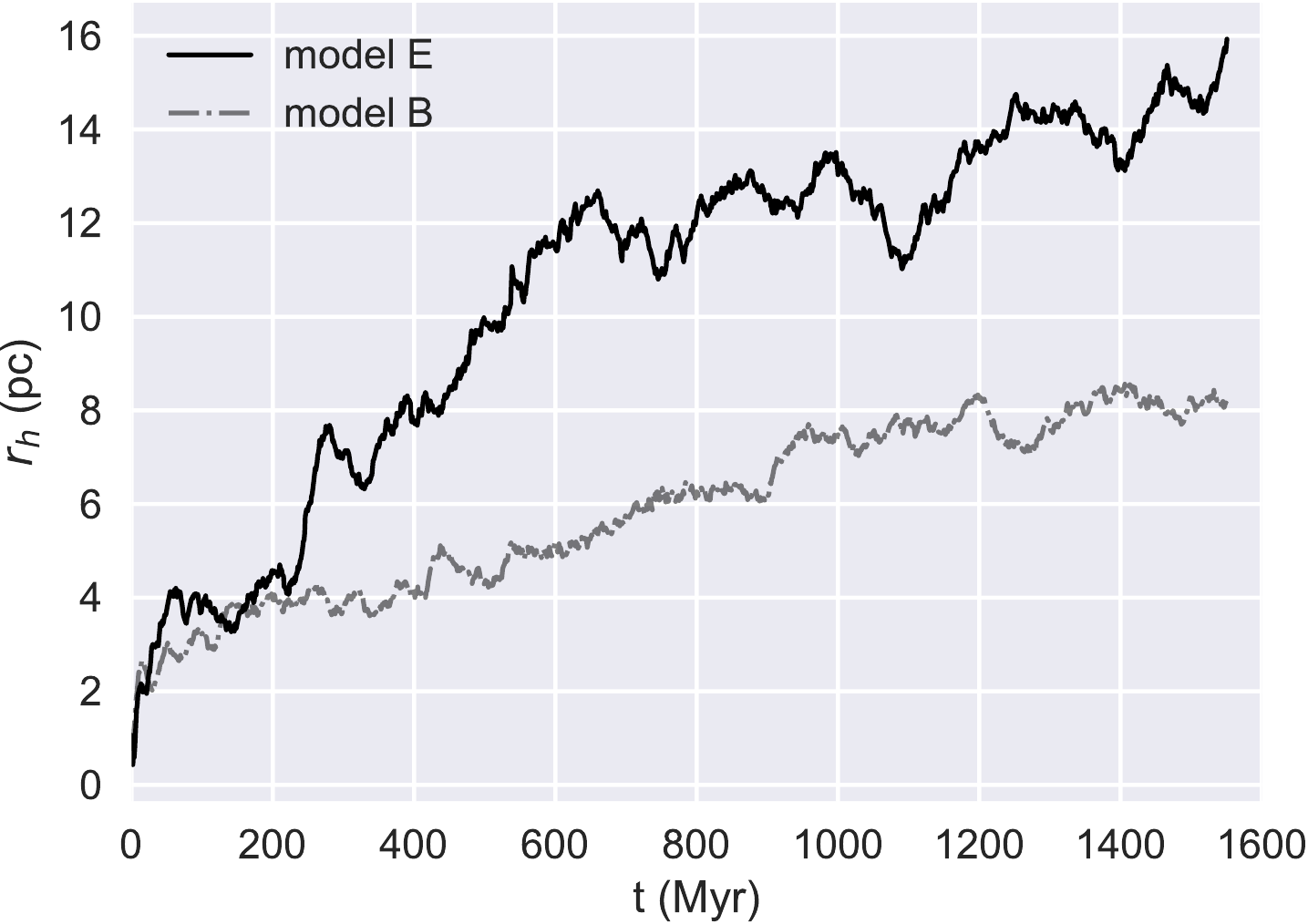}
\caption[Evolution of the cluster mass and half mass radius.]
{Evolution of the mass ($M_{cl}$, \textit{left panel}) and of the half mass radius
($r_{h}$, \textit{right panel}) for two of the simulations in the sets of model B (initial virial equilibrium)  and E (sub-virial).
}
\label{fig:trm}
\end{figure}

In Fig. \ref{fig:trm} we show the time evolution of the cluster mass ($M_{cl}$)
and of the half-mass radius ($r_{h}$) for two simulations of model B and E. We
notice, in both cases, a quick mass loss from the systems (few hundreds Myr)
followed by a secular trend. Model E (initially sub$-$virial) loses more mass
than model B. Both systems globally (right panel) expand, but after $\sim
200$ Myr, we notice that the half mass radius of the initially sub-virial model
shows a larger expansion.  This is a consequence of the initial violent
collapse of model E whose following relaxation determines its further, secular,
increased expansion and mass loss with respect to the virial case (model B)
(see also \cite{Terlevich,bt}).

\subsection{Cluster Velocity Dispersion}
\label{sigma}

\begin{figure}[htb]
    \begin{minipage}[t]{.3\textwidth}
        \centering
        \includegraphics[width=1.1\textwidth]{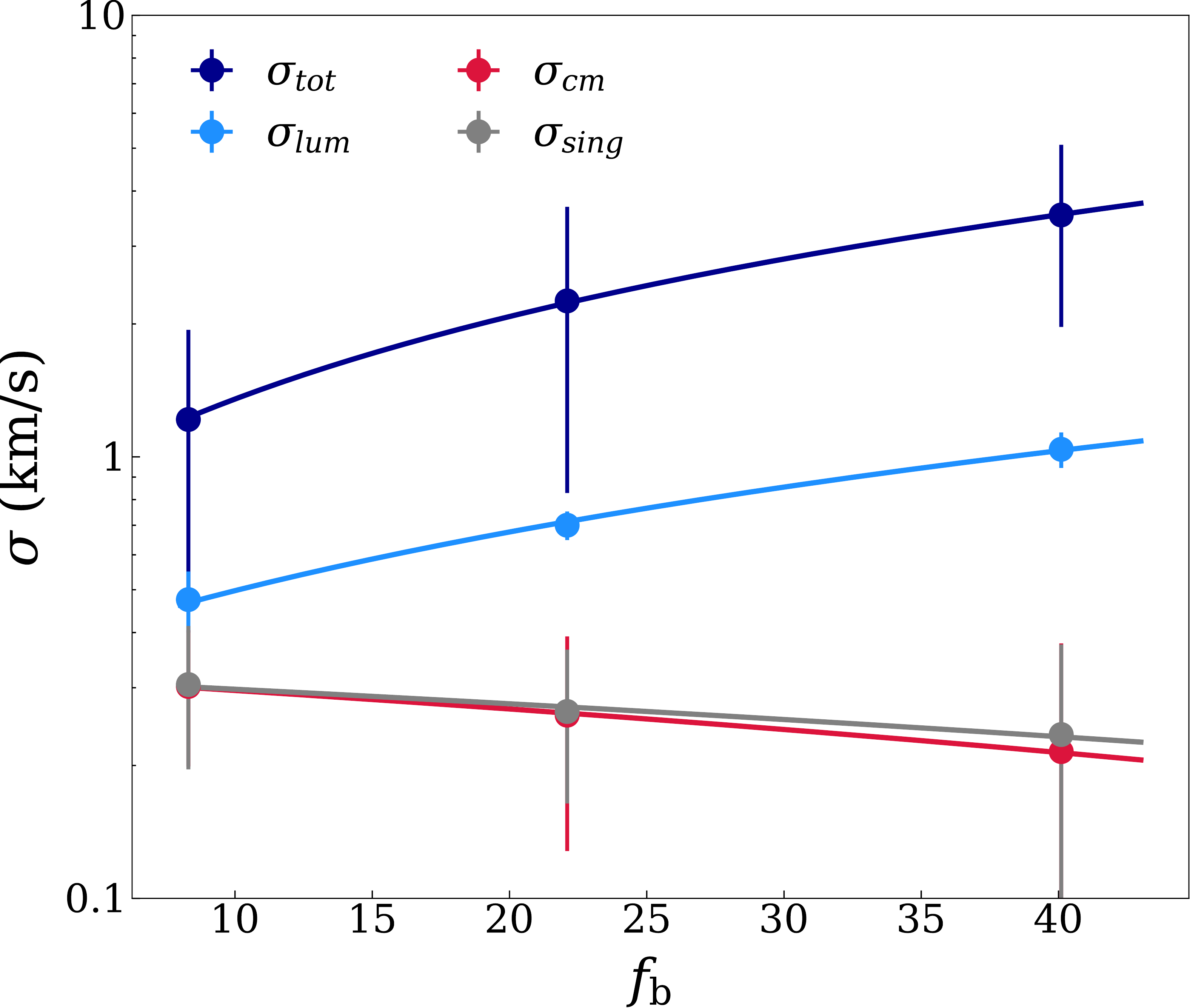}\quad
        {Models A, B and C, \textit{sample2}.}
    \end{minipage}
     \hfill
         \begin{minipage}[t]{.3\textwidth}
        \centering
        \includegraphics[width=1.1\textwidth]{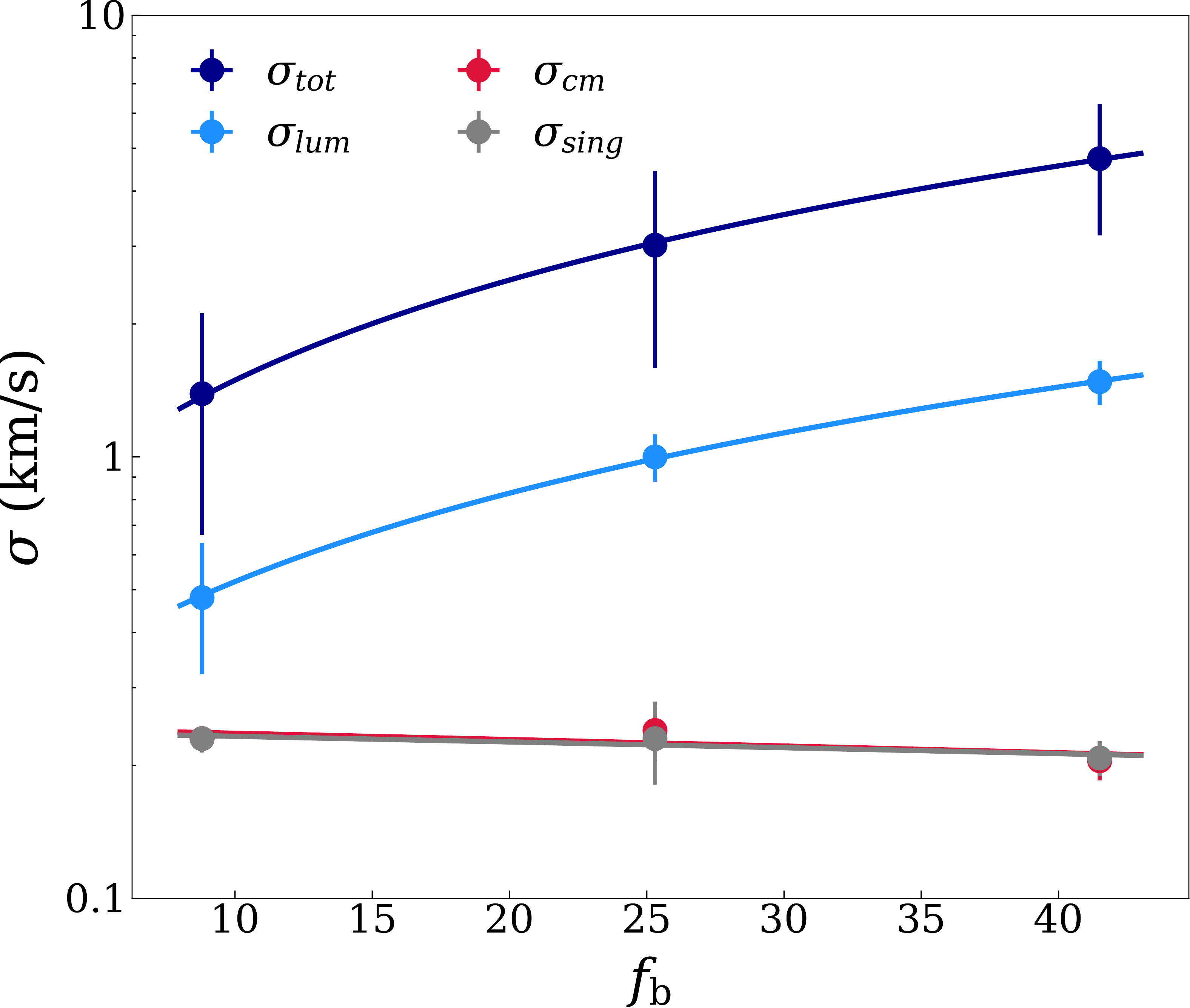}\quad
        {Models A, B and C, \textit{sample1}.}
    \end{minipage}
    \hfill
    \begin{minipage}[t]{.3\textwidth}
        \centering
        \includegraphics[width=1.1\textwidth]{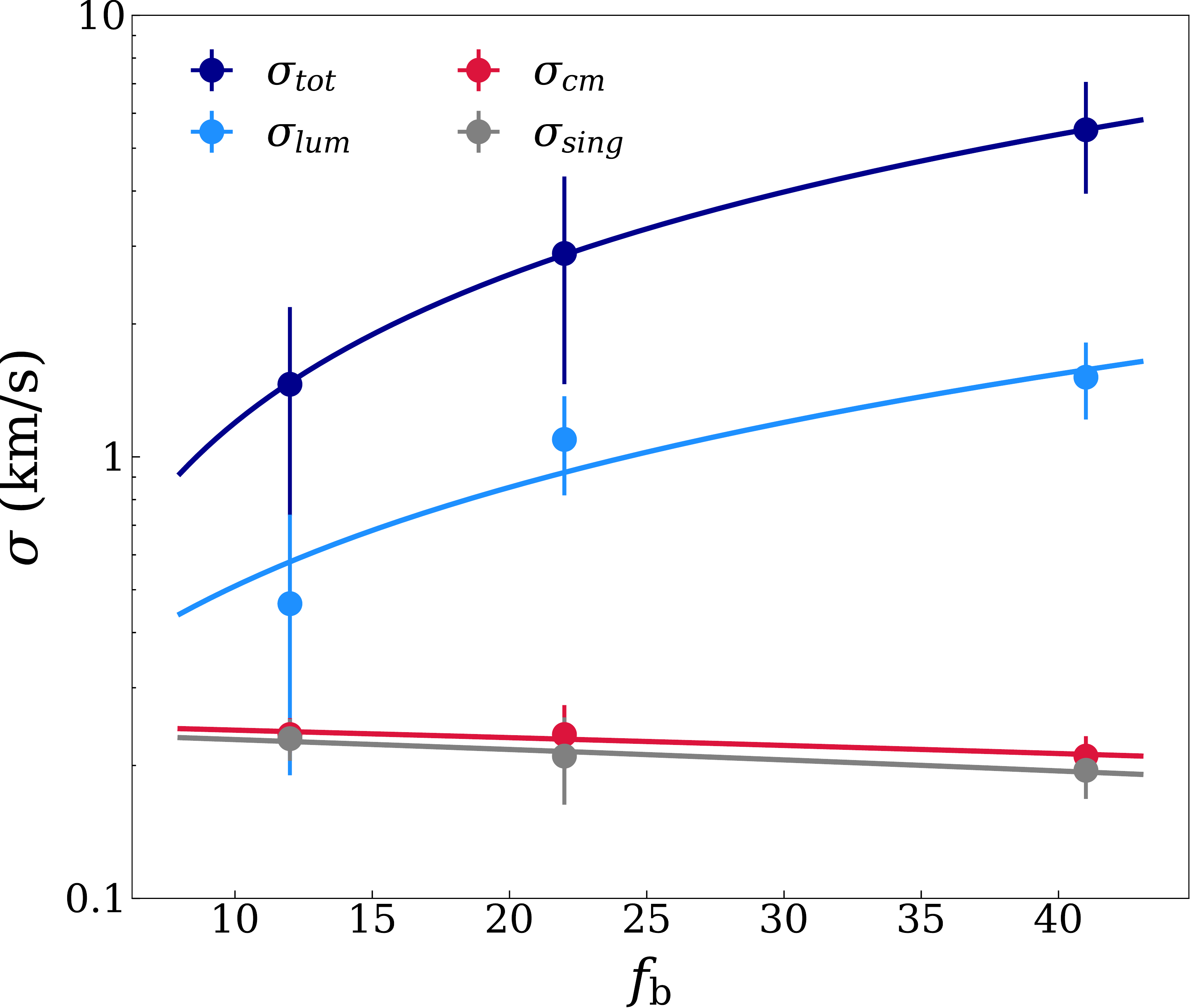}\quad
        {Models A, B and C, \textit{sample0}.}
    \end{minipage}
    \medskip
    \vspace{0.5cm}
    \begin{minipage}[t]{.3\textwidth}
        \centering
        \includegraphics[width=1.1\textwidth]{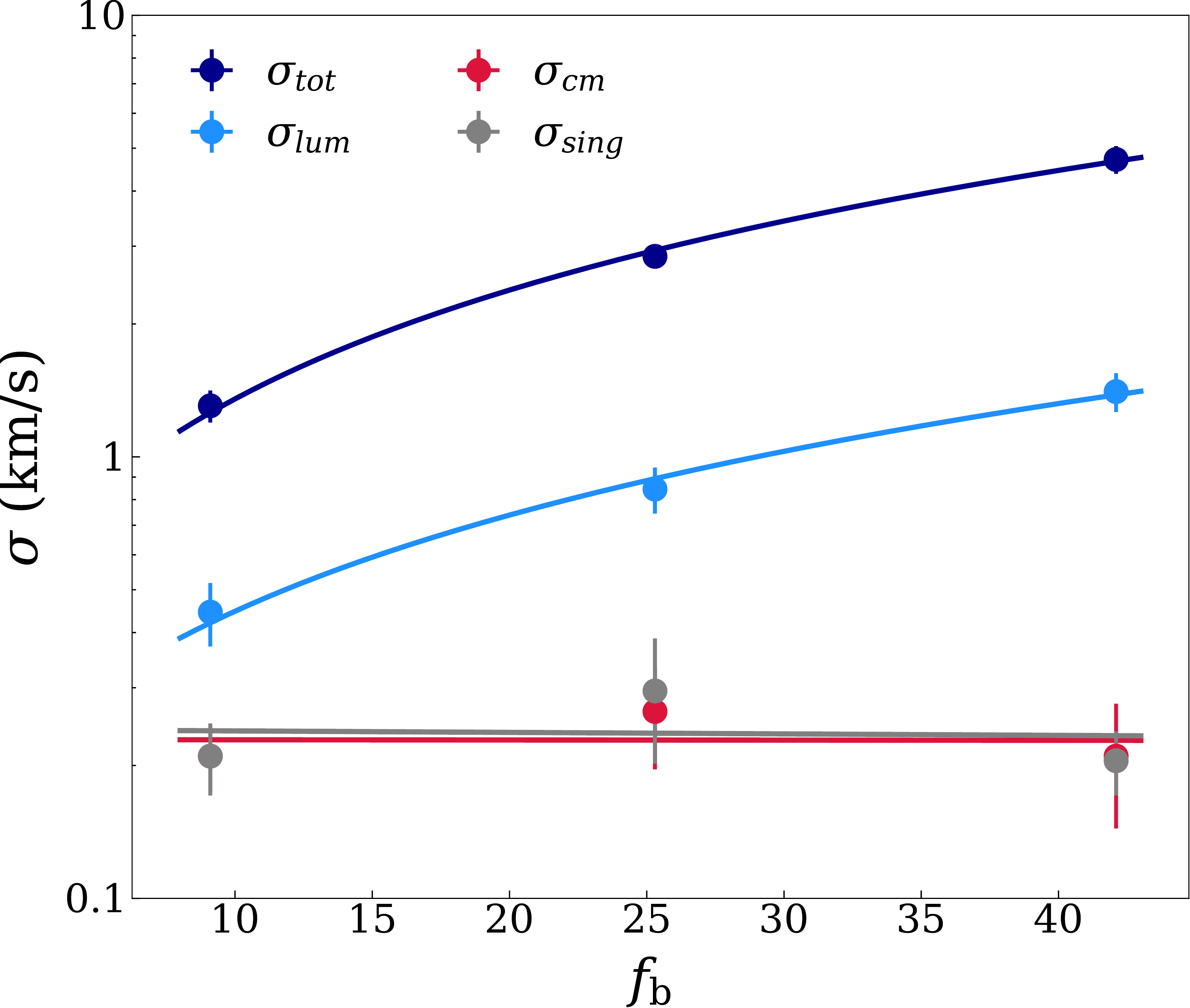}\quad
        {Models D, E and F, \textit{sample2}.}
    \end{minipage}
    \hfill
    \begin{minipage}[t]{.3\textwidth}
        \centering
        \includegraphics[width=1.1\textwidth]{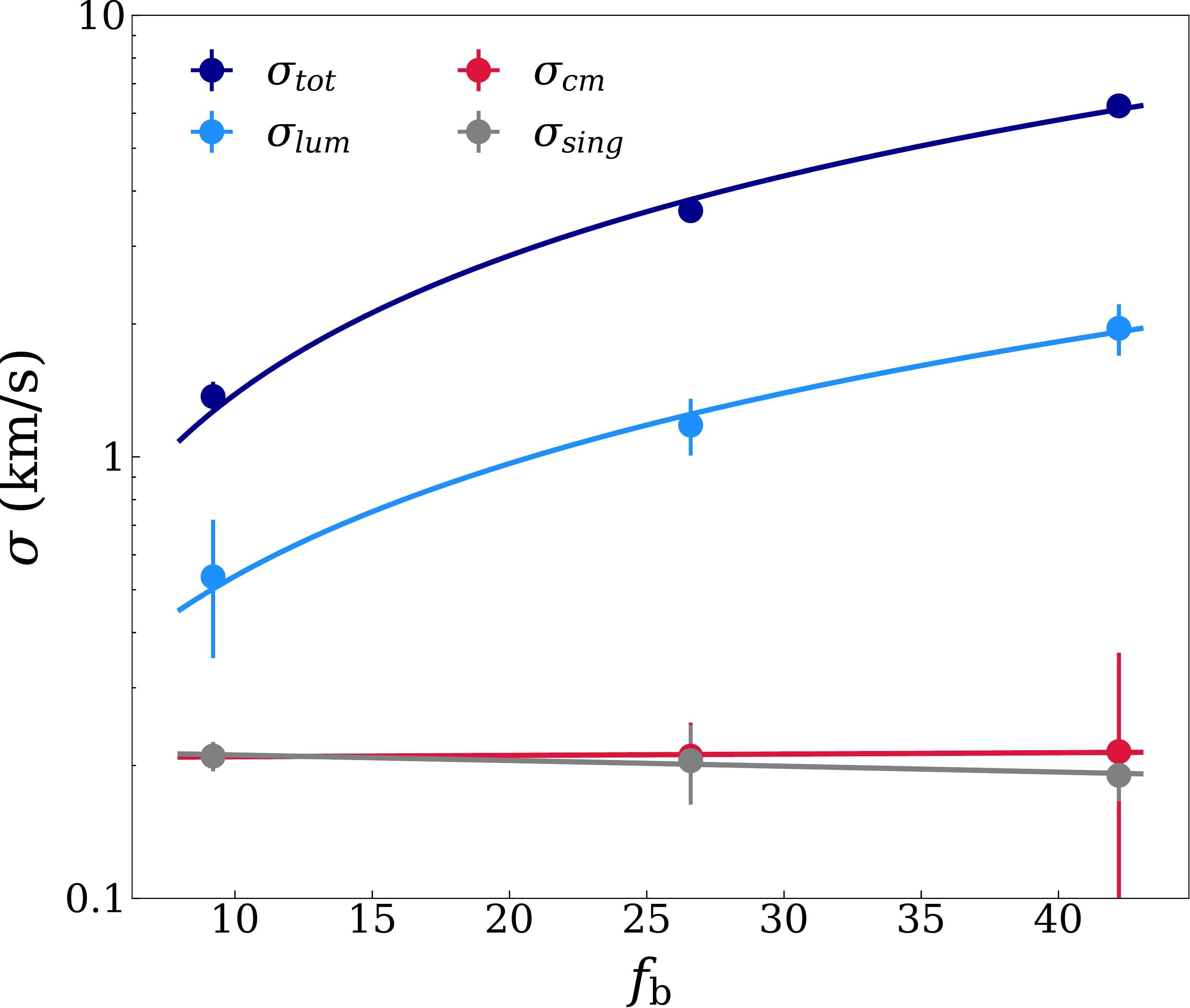}\quad
        {Models D, E and F, \textit{sample1}.}
    \end{minipage}
    \hfill
    \begin{minipage}[t]{.3\textwidth}
        \centering
        \includegraphics[width=1.1\textwidth]{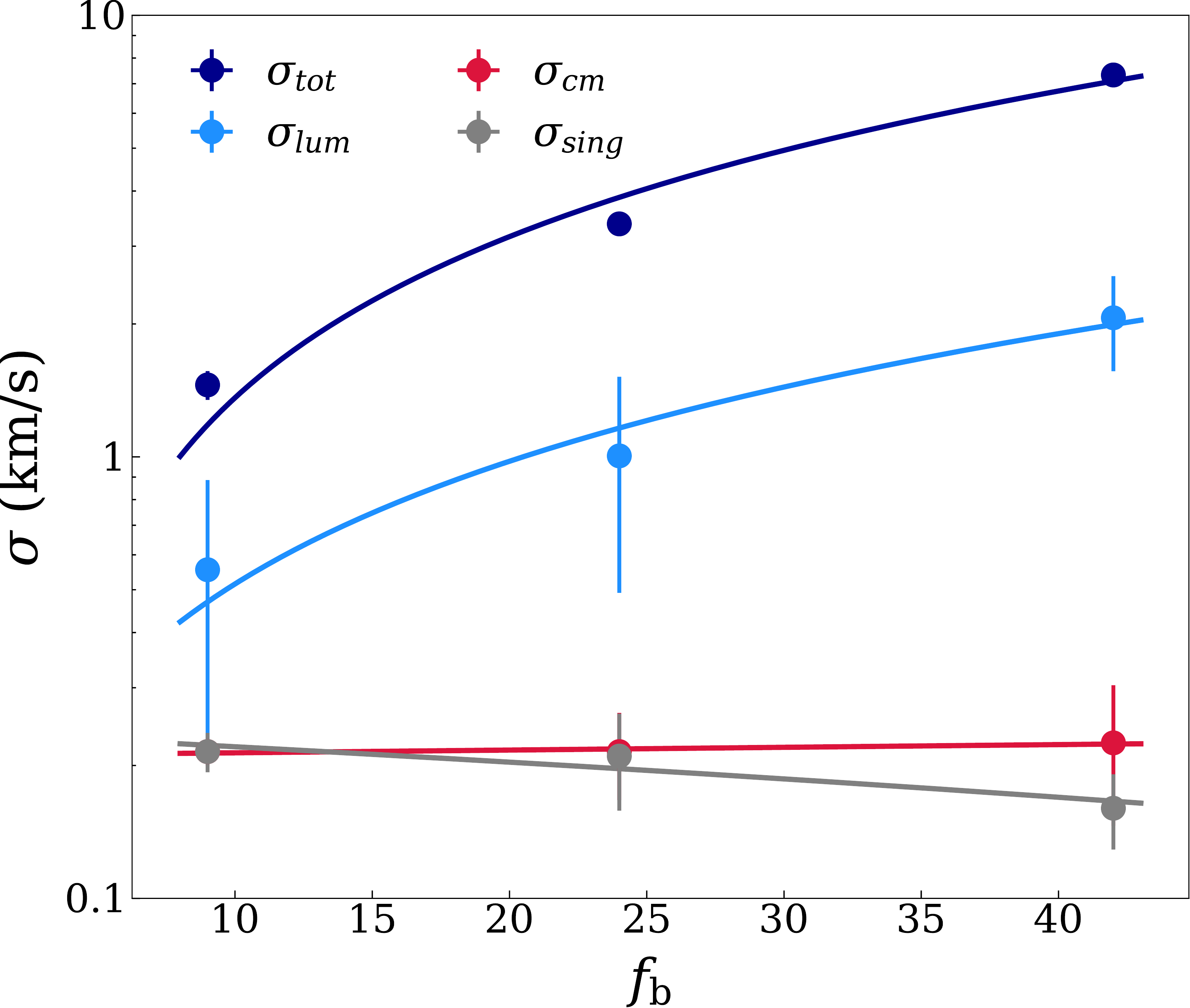}\quad
        {Models D, E and F, \textit{sample0}.}
    \end{minipage}
    \vspace{0.5cm}
    \caption{Velocity dispersions estimated according to the four methods described
in Sect. \ref{mm} versus the percentage of binaries  at $1.5$ Gyr $\pm 50$ Myr (see Table \ref{frac_tot})
for the models A, B, and C (\textit{top panels}), and  D, E and F (\textit{bottom panels}) for
 \textit{sample2}, \textit{sample1} and \textit{sample0} respectively from left to right.}
\label{fig:six}
\end{figure}

As discussed in Sect. \ref{stra} (see Fig. \ref{fig:hrD})  we have extracted
three samples of cluster stars from each model basing on different luminosity
cut-off.  We estimate the velocity dispersion for each sample by means of the
four methods described in Sect. \ref{stra}. Figure \ref{fig:six} shows the
averaged cluster velocity dispersion at $1.5$ Gyr as function of the binary
fraction for \textit{sample2}, \textit{sample1}, and \textit{sample0}. The
values of the velocity dispersion (averaged over a time$-$range of about $100$
Myr) are means over the whole set of the 10 $N$-body simulations performed for each
model. A summary of the results is given in Table \ref{table:sigma_2},
\ref{table:sigma_1} and \ref{table:sigma_0} for \textit{sample2},
\textit{sample1} and \textit{sample0}, respectively.  In Table \ref{frac_tot},
we summarize the properties of the clusters (number of stars, mass, and
percentage of binaries) averaged for each model at $1.5$ Gyr for the three
samples.

Since binaries are usually more luminous (and also more massive, so
less likely to escape) than single stars their contribution to the sample
results larger than the contribution of individual stars. Such effect is thus
reflected in the percentage of binary at that time.

Considering \textit{sample2}, from Fig.\ref{fig:six} (first column, top panel)
and Table \ref{table:sigma_2}, we note the, expected, result that the velocity
dispersion estimated with method 1 ($\sigma_{tot}$) is significantly larger,
being in the range between $1.215$ km/s $\leq \sigma_{tot} \leq 4.715 $ km/s,
with respect to the velocity dispersion derived with the other methods.
Actually, in this case, the binary orbital motion inflates the estimate of the
velocity dispersion with respect to the global orbital motion. On the other
hand, when evaluating the velocity dispersion with method 3, thus weighing the
binary contribution with the luminosity of the components, the result ranges
between $0.445$  $\leq \sigma_{lum}$ (km/s) $\leq  1.405 $. Both $\sigma_{tot}$
and $\sigma_{lum}$ increase as the fraction of binaries in the sample
increases.

On the contrary, the velocity  dispersion derived with method 2 ($\sigma_{cm}$)
is independent of the binary content in the sample as shown in Fig.
\ref{fig:six}.  In fact, it does not show any trend and correlation with the
fraction of binaries in the sample. The velocity dispersion obtained with such
method is much smaller, ranging between $0.2 $ km/s $ \leq \sigma_{cm} \leq 0.3
$ km/s, than when considering all stars individually.

We also estimated the velocity dispersion excluding binary stars (method 4):
the velocity dispersion estimated in this way, $\sigma_{sing}$, is very similar
to $\sigma_{cm}$ for all the samples.

Similar results are found for \textit{sample1} (Fig. \ref{fig:six}, second
column and Table \ref{table:sigma_1}) and for \textit{sample0} (Fig.
\ref{fig:six}, third column and Table \ref{table:sigma_0}). As the luminosity
threshold increases (from \textit{sample2} to \textit{sample0}) the velocity
dispersion estimated with $\sigma_{tot}$ and $\sigma_{lum}$ increases too.
This outcome is coherent with the fact that such methods are binary-dependent
(the more luminous the sample, the larger the fraction of binaries).

If we compare the models studied, see top and bottom rows of Fig.
\ref{fig:six}, we notice that the ``virial'' models A, B and C show generally
smaller values of the velocity dispersion with respect to ``sub-virial'' models
D, E and F. Such outcome is common to all the three samples and it is evident
for $\sigma_{tot}$ and $\sigma_{lum}$. This effect is  a combined consequence
of the different fraction of binary stars between models on an initial virial
and sub$-$virial equilibrium as described in Table \ref{frac_tot}. On the other
hand, the velocity dispersion $\sigma_{cm}$ and that obtained from single
stars, $\sigma_{sing}$, match very well each other, over the three samples.

\begin{table}[!h]
\centering
\begin{tabular}{@{}ccccc@{}}
\toprule
\multicolumn{5}{c}{\textbf{\textit{Sample2}}}                                                                                                                    \\ \midrule
\textbf{Model} & \textbf{$\sigma_{tot}$} & \textbf{$\sigma_{cm}$} & \textbf{$\sigma_{lum}$} & \textbf{$\sigma_{sing}$} \\
               \\ \midrule
               &                         &                        & \multicolumn{1}{l}{}                        & \multicolumn{1}{l}{}                           \\
A              &    1.215    $\pm$   0.109   &    0.302    $\pm$   0.063   &    0.475    $\pm$   0.075   &    0.305    $\pm$   0.069                            \\
               &                         &                        &                                             &                                                \\
B              &    2.255    $\pm$   0.106   &    0.260    $\pm$   0.032   &    0.700    $\pm$   0.052   &    0.265    $\pm$   0.041                            \\
               &                         &                        &                                             &                                                \\
C              &    3.530    $\pm$   0.341   &    0.215    $\pm$   0.063   &    1.040    $\pm$   0.096   &    0.235    $\pm$   0.141                            \\
               &                         &                        &                                             &                                                \\
D              &    1.305    $\pm$   0.117   &    0.210    $\pm$   0.036   &    0.445    $\pm$   0.073   &    0.210    $\pm$   0.039                            \\
               &                         &                        &                                             &                                                \\
E              &    2.845    $\pm$   0.266   &    0.265    $\pm$   0.069   &    0.845    $\pm$   0.101   &    0.295    $\pm$   0.093                            \\
               &                         &                        &                                             &                                                \\
F              &    4.715    $\pm$   0.523   &    0.210    $\pm$   0.066   &    1.405    $\pm$   0.142   &    0.205    $\pm$   0.034                            \\
               &                         &                        &                                             &                                                \\ \bottomrule
\end{tabular}
\caption{The velocity dispersions (in km/s) obtained with each method described in Sect. \ref{stra}, for \textit{sample2}.
The columns indicate, from left to
right: 1) the model ID, 2) the velocity dispersion obtained with
method 1 ($\sigma_{tot}$), 3) the velocity dispersion obtained with method 2 ($\sigma_{cm}$),
4) the velocity dispersion obtained with method 3 ($\sigma_{lum}$), 5) the velocity
dispersion obtained with method 4 ($\sigma_{sing}$).
The error of the velocity dispersion is evaluated
according to its standard deviation measures.}
\label{table:sigma_2}
\end{table}

\begin{table}[!h]
\centering
\begin{tabular}{@{}ccccc@{}}
\toprule
\multicolumn{5}{c}{\textbf{\textit{Sample1}}}                                                                                                                    \\ \midrule
\textbf{Model} & \textbf{$\sigma_{tot}$} & \textbf{$\sigma_{cm}$} & \textbf{$\sigma_{lum}$} & \textbf{$\sigma_{sing}$} \\ \midrule
               &                         &                        & \multicolumn{1}{l}{}                        & \multicolumn{1}{l}{}                           \\
A              &    1.390    $\pm$   0.227   &    0.230    $\pm$   0.013   &    0.480    $\pm$   0.158   &    0.230    $\pm$   0.016                                  \\
               &                         &                        &                                             &                                                \\
B              &    3.015    $\pm$   0.424   &    0.240    $\pm$   0.034   &    1.000    $\pm$   0.124   &    0.230    $\pm$   0.049                                   \\
               &                         &                        &                                             &                                                \\
C              &    4.735    $\pm$   0.585   &    0.205    $\pm$   0.020   &    1.480    $\pm$   0.170   &    0.208    $\pm$   0.019                                   \\
               &                         &                        &                                             &                                                \\
D              &    1.370    $\pm$   0.303   &    0.210    $\pm$   0.012   &    0.535    $\pm$   0.185   &    0.210    $\pm$   0.016                                   \\
               &                         &                        &                                             &                                                \\
E              &    3.610    $\pm$   0.741   &    0.210    $\pm$   0.040   &    1.180    $\pm$   0.173   &    0.205    $\pm$   0.042                                   \\
               &                         &                        &                                             &                                                \\
F              &    6.235    $\pm$   1.003   &    0.215    $\pm$   0.145   &    1.955    $\pm$   0.260   &    0.190    $\pm$   0.024                                   \\
               &                         &                        &                                             &                                                \\ \bottomrule
\end{tabular}
\caption{Same as in Table \ref{table:sigma_2} but for \textit{sample1}.}
\label{table:sigma_1}
\end{table}

\begin{table}[!h]
\centering
\begin{tabular}{@{}ccccc@{}}
\toprule
\multicolumn{5}{c}{\textbf{\textit{Sample0}}}                                                                                                                    \\ \midrule
\textbf{Model} & \textbf{$\sigma_{tot}$} & \textbf{$\sigma_{cm}$} & \textbf{$\sigma_{lum}$} & \textbf{$\sigma_{sing}$} \\
               \midrule
               &                         &                        & \multicolumn{1}{l}{}                        & \multicolumn{1}{l}{}                           \\
A              &    1.460    $\pm$   0.561   &    0.235    $\pm$   0.021   &    0.465    $\pm$   0.275   &    0.230    $\pm$   0.025                                    \\
               &                         &                        &                                             &                                                \\
B              &    2.888    $\pm$   0.558   &    0.235    $\pm$   0.039   &    1.095    $\pm$   0.277   &    0.210    $\pm$   0.047                                    \\
               &                         &                        &                                             &                                                \\
C              &    5.505    $\pm$   1.082   &    0.210    $\pm$   0.023   &    1.515    $\pm$   0.300   &    0.195    $\pm$   0.027                                      \\
               &                         &                        &                                             &                                                \\
D              &    1.455    $\pm$   0.724   &    0.215    $\pm$   0.015   &    0.555    $\pm$   0.331   &    0.215    $\pm$   0.022                                     \\
               &                         &                        &                                             &                                                \\
E              &    3.370    $\pm$   1.427   &    0.215    $\pm$   0.048   &    1.005    $\pm$   0.513   &    0.210    $\pm$   0.052                                     \\
               &                         &                        &                                             &                                                \\
F              &    7.325    $\pm$   1.561   &    0.225    $\pm$   0.079   &    2.065    $\pm$   0.502   &    0.160    $\pm$   0.031                                      \\
               &                         &                        &                                             &                                                \\ \bottomrule
\end{tabular}
\caption[Same as in Table \ref{table:sigma_2} but for \textit{sample0}.]
{Same as in Table \ref{table:sigma_2} but for \textit{sample0}.}
\label{table:sigma_0}
\end{table}

\begin{table}[!h]
\centering
\begin{tabular}{cccccccccc}
\hline
\multicolumn{1}{l}{}                        & \multicolumn{3}{c}{\multirow{2}{*}{\textit{Sample2}}}                                      & \multicolumn{3}{c}{\multirow{2}{*}{\textit{Sample1}}}                                      & \multicolumn{3}{c}{\multirow{2}{*}{\textit{Sample0}}}                 \\
\multicolumn{1}{l}{}                        & \multicolumn{3}{c}{}                                                              & \multicolumn{3}{c}{}                                                              & \multicolumn{3}{c}{}                                         \\ \hline
\multicolumn{1}{c|}{\multirow{2}{*}{Model}} & \multirow{2}{*}{$\langle N \rangle$} & \multirow{2}{*}{$\langle M_{cl} \rangle $ (M$_{\odot}$)} & \multicolumn{1}{c|}{\multirow{2}{*}{$\langle f_b \, \rangle\ (\%) $}} & \multirow{2}{*}{$\langle N \rangle$} & \multirow{2}{*}{$\langle M_{cl} \rangle $ (M$_{\odot}$)} & \multicolumn{1}{c|}{\multirow{2}{*}{$\langle f_b \,  \rangle\ (\%) $}} & \multirow{2}{*}{$\langle N \rangle$} & \multirow{2}{*}{$\langle M_{cl} \rangle $ (M$_{\odot}$)} & \multirow{2}{*}{$\langle f_b \,  \rangle \ (\%)$} \\
\multicolumn{1}{c|}{}                       &                    &                    & \multicolumn{1}{c|}{}                   &                    &                    & \multicolumn{1}{c|}{}                   &                    &                    &                    \\ \hline
\multicolumn{1}{c|}{}                       &                    &                    & \multicolumn{1}{c|}{}                   &                    &                    & \multicolumn{1}{c|}{}                   &                    &                    &                    \\
\multicolumn{1}{c|}{A}                      & 392             & 277             & \multicolumn{1}{c|}{8.3}               & 136             & 156               & \multicolumn{1}{c|}{8.8}               & 50              & 81             & 12.0              \\
\multicolumn{1}{c|}{}                       &                    &                    & \multicolumn{1}{c|}{}                   &                    &                    & \multicolumn{1}{c|}{}                   &                    &                    &                    \\
\multicolumn{1}{c|}{B}                      & 441             & 266             & \multicolumn{1}{c|}{22.1}              & 138             & 137             & \multicolumn{1}{c|}{25.3}              & 54              & 73             & 22.3              \\
\multicolumn{1}{c|}{}                       &                    &                    & \multicolumn{1}{c|}{}                   &                    &                    & \multicolumn{1}{c|}{}                   &                    &                    &                    \\
\multicolumn{1}{c|}{C}                      & 507             & 278             & \multicolumn{1}{c|}{40.1}              & 168             & 147             & \multicolumn{1}{c|}{41.5}              & 68              & 74             & 41.0              \\
\multicolumn{1}{l|}{}                       & \multicolumn{1}{l}{} & \multicolumn{1}{l}{} & \multicolumn{1}{l|}{}                   & \multicolumn{1}{l}{} & \multicolumn{1}{l}{} & \multicolumn{1}{l|}{}                   & \multicolumn{1}{l}{} & \multicolumn{1}{l}{} & \multicolumn{1}{l}{} \\ \hline
\multicolumn{1}{c|}{}                       &                    &                    & \multicolumn{1}{c|}{}                   &                    &                    & \multicolumn{1}{c|}{}                   &                    &                    &                    \\
\multicolumn{1}{c|}{D}                      & 352             & 251             & \multicolumn{1}{c|}{9.1}               & 109             & 134             & \multicolumn{1}{c|}{9.2}               & 47              & 75             & 9.0              \\
\multicolumn{1}{c|}{}                       &                    &                    & \multicolumn{1}{c|}{}                   &                    &                    & \multicolumn{1}{c|}{}                   &                    &                    &                    \\
\multicolumn{1}{c|}{E}                      & 377             & 251             & \multicolumn{1}{c|}{25.3}              & 124             & 135             & \multicolumn{1}{c|}{26.6}              & 50              & 66             & 24.1              \\
\multicolumn{1}{c|}{}                       &                    &                    & \multicolumn{1}{c|}{}                   &                    &                    & \multicolumn{1}{c|}{}                   &                    &                    &                    \\
\multicolumn{1}{c|}{F}                      & 444             & 275             & \multicolumn{1}{c|}{42.1}              & 155             & 141             & \multicolumn{1}{c|}{42.2}              & 67              & 81             & 42.1              \\
\multicolumn{1}{c|}{}                       &                    &                    & \multicolumn{1}{c|}{}                   &                    &                    & \multicolumn{1}{c|}{}                   &                    &                    &                    \\ \hline
\end{tabular}
\caption{Parameters characterizing the cluster stellar population for the three samples studied, at $t=1.5$ Gyr.
The columns represent (from left to the right): 1) the model ID, 2) the total number of star ($N$), 3) the mass of the cluster ($M_{cl}$), 4) the percentage of binary stars ($f_b$).
The reported values are averaged over all the simulations performed for each model.
}

\label{frac_tot}
\end{table}

\subsection{Time evolution of the velocity dispersion}
\label{global_ev}

\begin{figure}[htb]
    \begin{minipage}[t]{.3\textwidth}
        \centering
        \includegraphics[width=\textwidth]{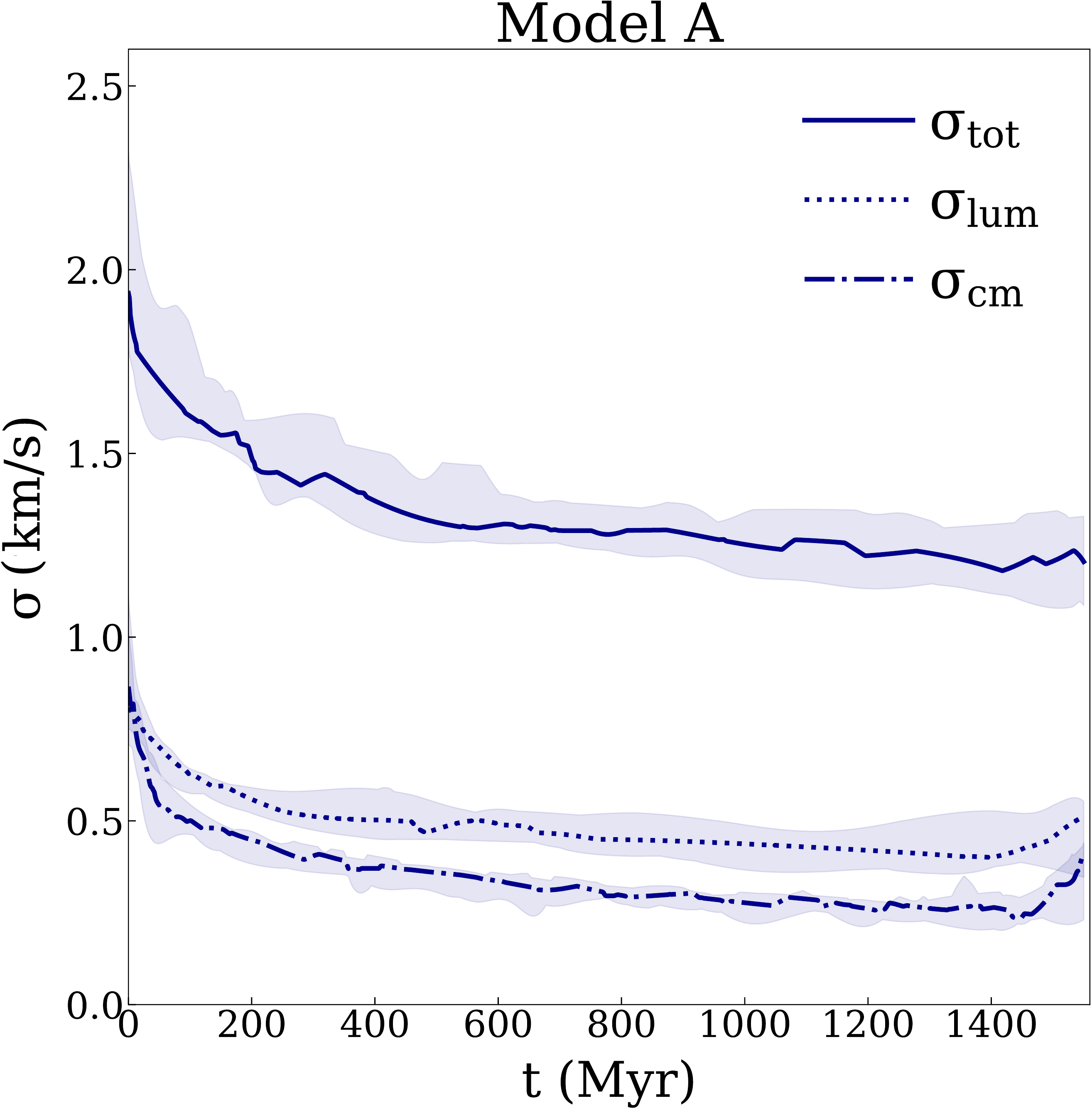}\quad
    \end{minipage}
    \hfill
    \begin{minipage}[t]{.3\textwidth}
        \centering
        \includegraphics[width=\textwidth]{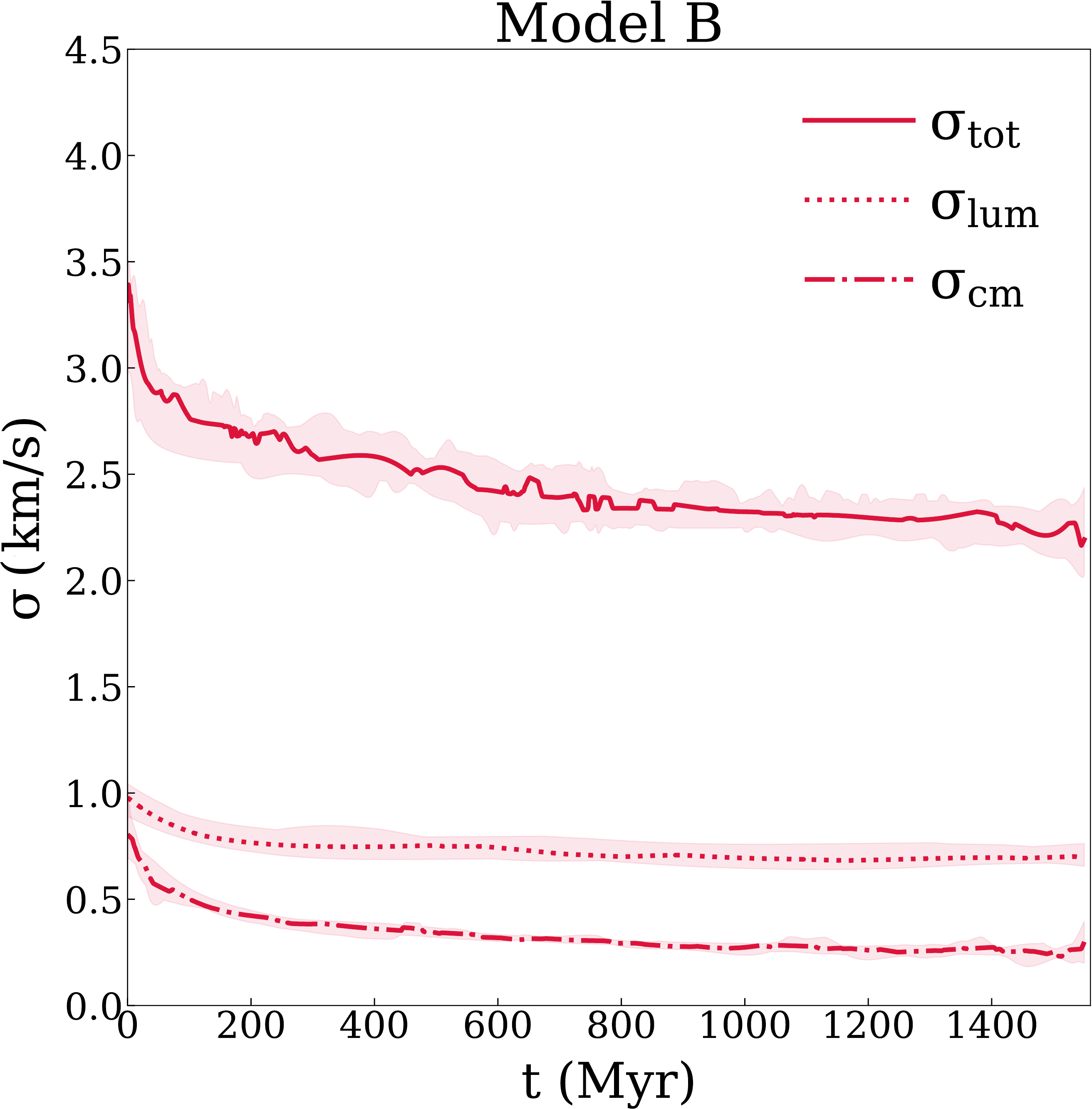}\quad
    \end{minipage}
    \hfill
    \begin{minipage}[t]{.29\textwidth}
        \centering
        \includegraphics[width=\textwidth]{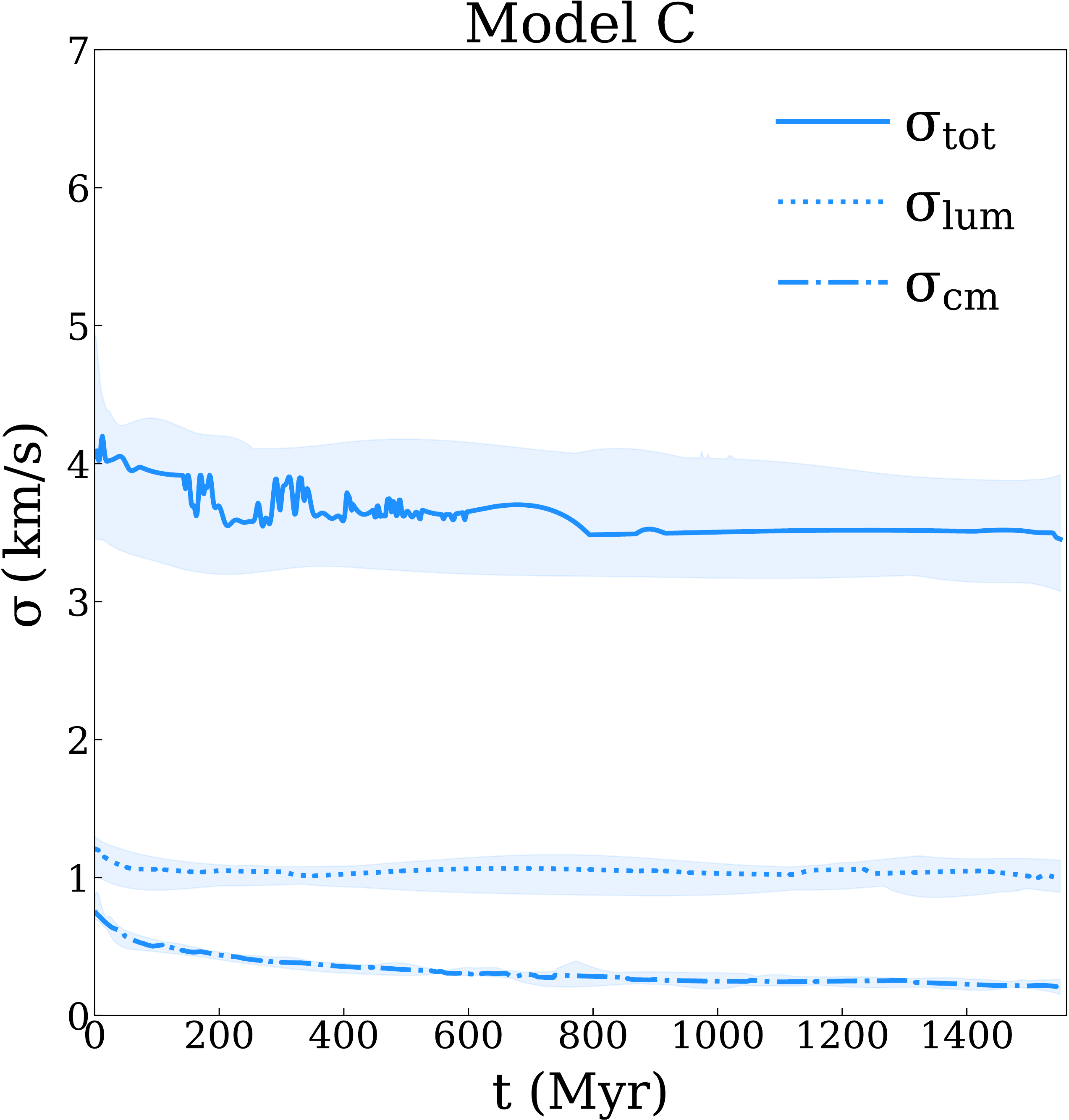}\quad
    \end{minipage}
    \medskip
    \vspace{0.5cm}
    \begin{minipage}[t]{.3\textwidth}
        \centering
        \includegraphics[width=\textwidth]{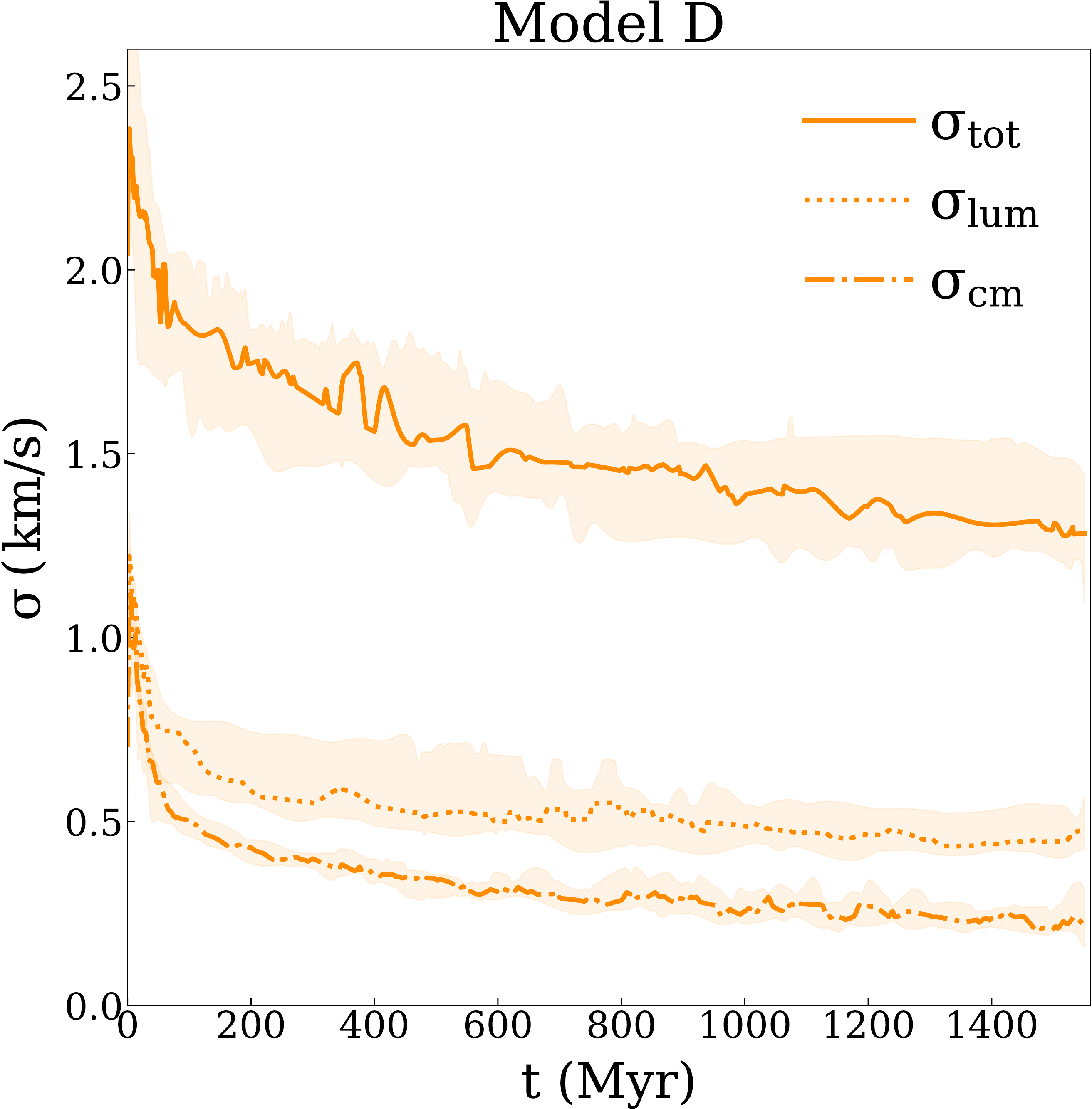}\quad
    \end{minipage}
    \hfill
    \begin{minipage}[t]{.3\textwidth}
        \centering
        \includegraphics[width=\textwidth]{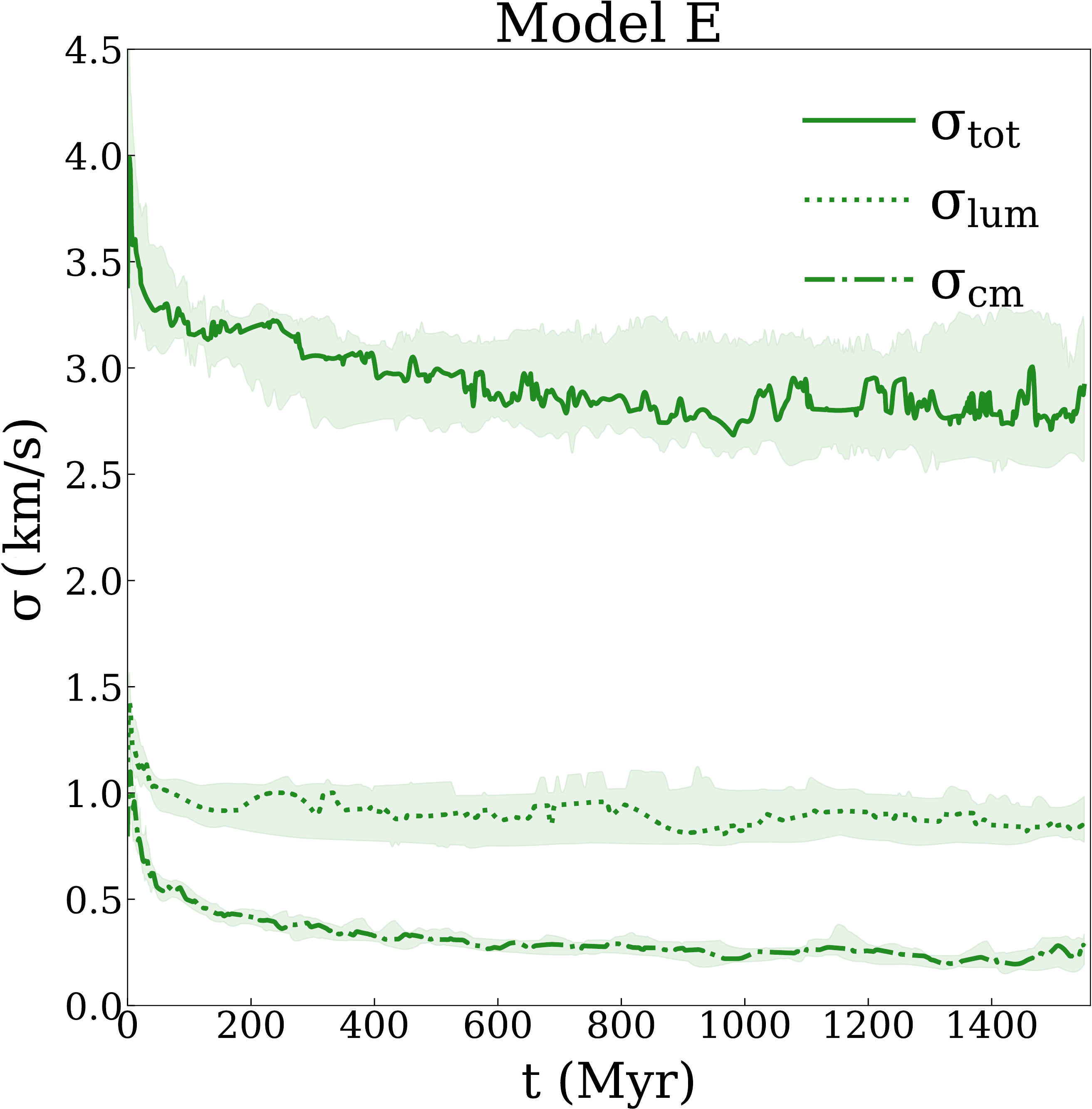}\quad
    \end{minipage}
    \hfill
    \begin{minipage}[t]{.29\textwidth}
        \centering
        \includegraphics[width=\textwidth]{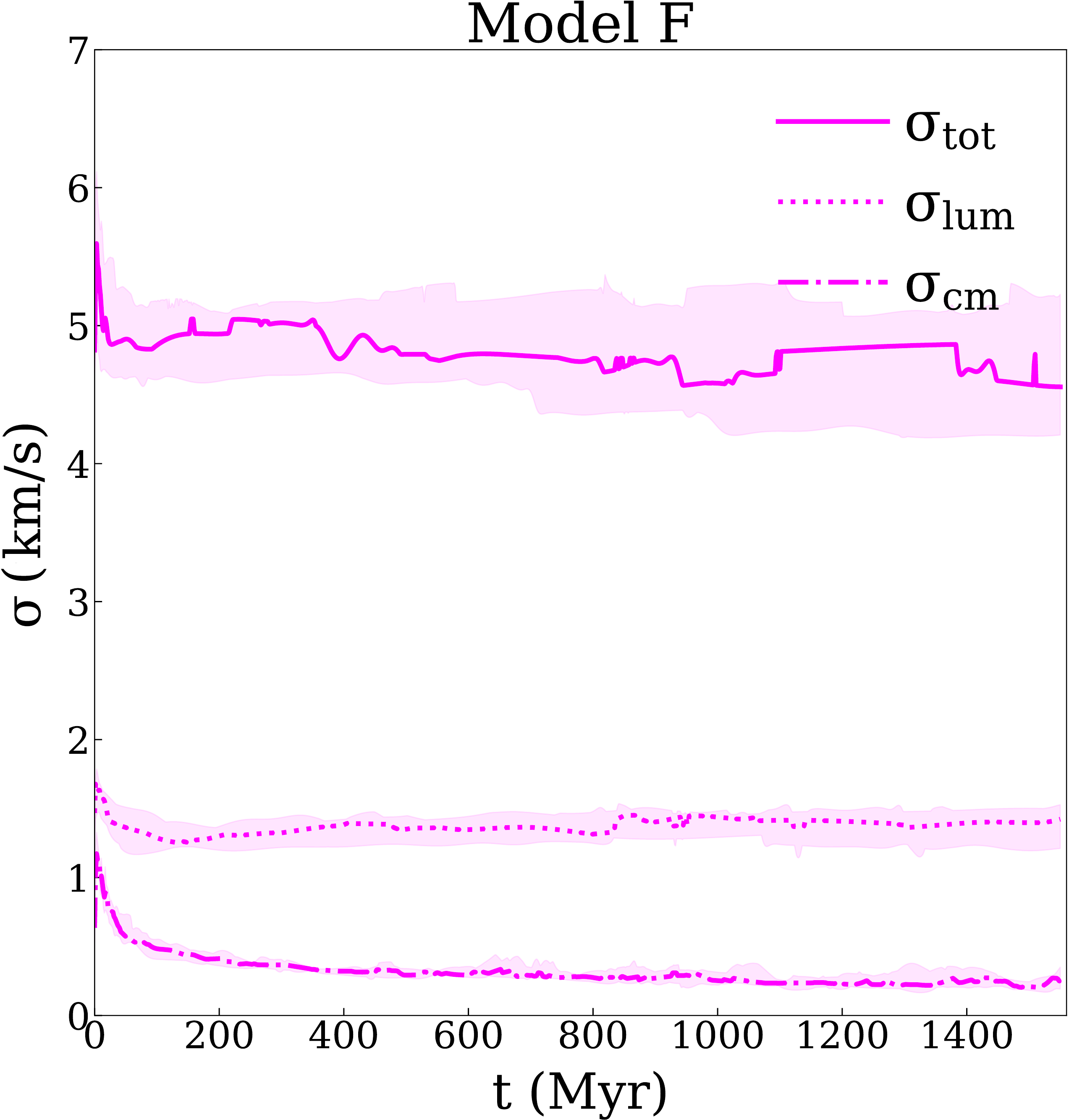}\quad
    \end{minipage}
    \hfill
    \caption{Time evolution of the various velocity dispersions for each model of \textit{sample2}.
    The different lines represent the different method used to estimate $\sigma$: $\sigma_{tot}$ solid line,
$\sigma_{lum}$ dotted line and $\sigma_{cm}$ dot-dashed line.  The shaded region on each line indicates
the error which is estimate as the standard deviation of the measures of $\sigma$. For display clarity, the results of $\sigma_{sing}$
are not plotted here (see Fig. \ref{fig:zoom}).}
\label{fig:allsigmaev}
\end{figure}

In Fig. \ref{fig:allsigmaev} we show the evolution of the velocity dispersion
averaged over all the simulations of each of the six models studied (A, B and
C: top panels, left to right; C, D and E: bottom panels, left to right) for \textit{sample2}. We
indicate with different line-style the velocity dispersion estimated with the
four methods explained in Sect. \ref{stra}.

As expected, $\sigma_{tot}$ is at
any time larger than the velocity dispersion estimated with the other methods.

\begin{figure}[!h]
\centering
\includegraphics[width=0.45\textwidth]{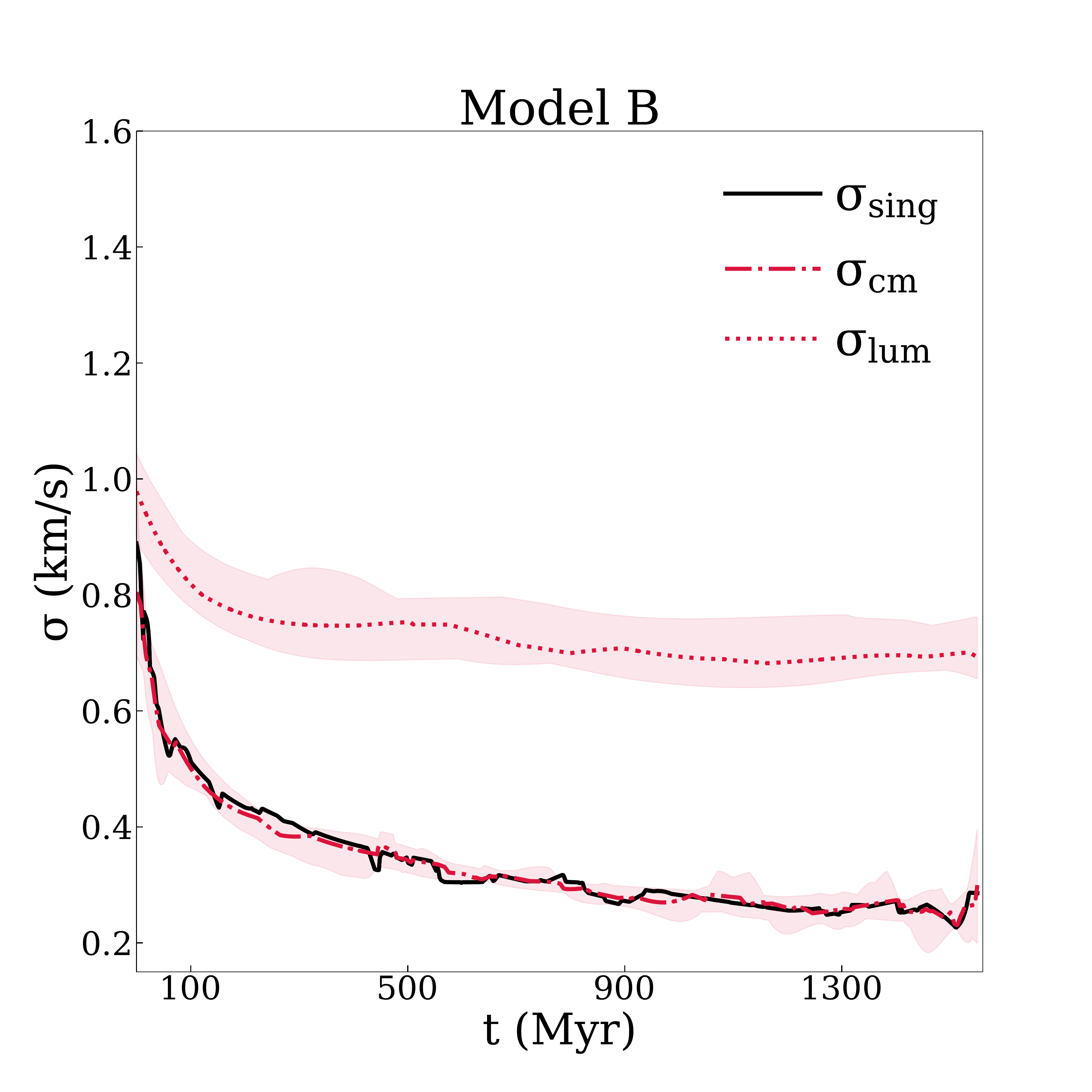}
\hspace{0.2cm}
\includegraphics[width=0.45\textwidth]{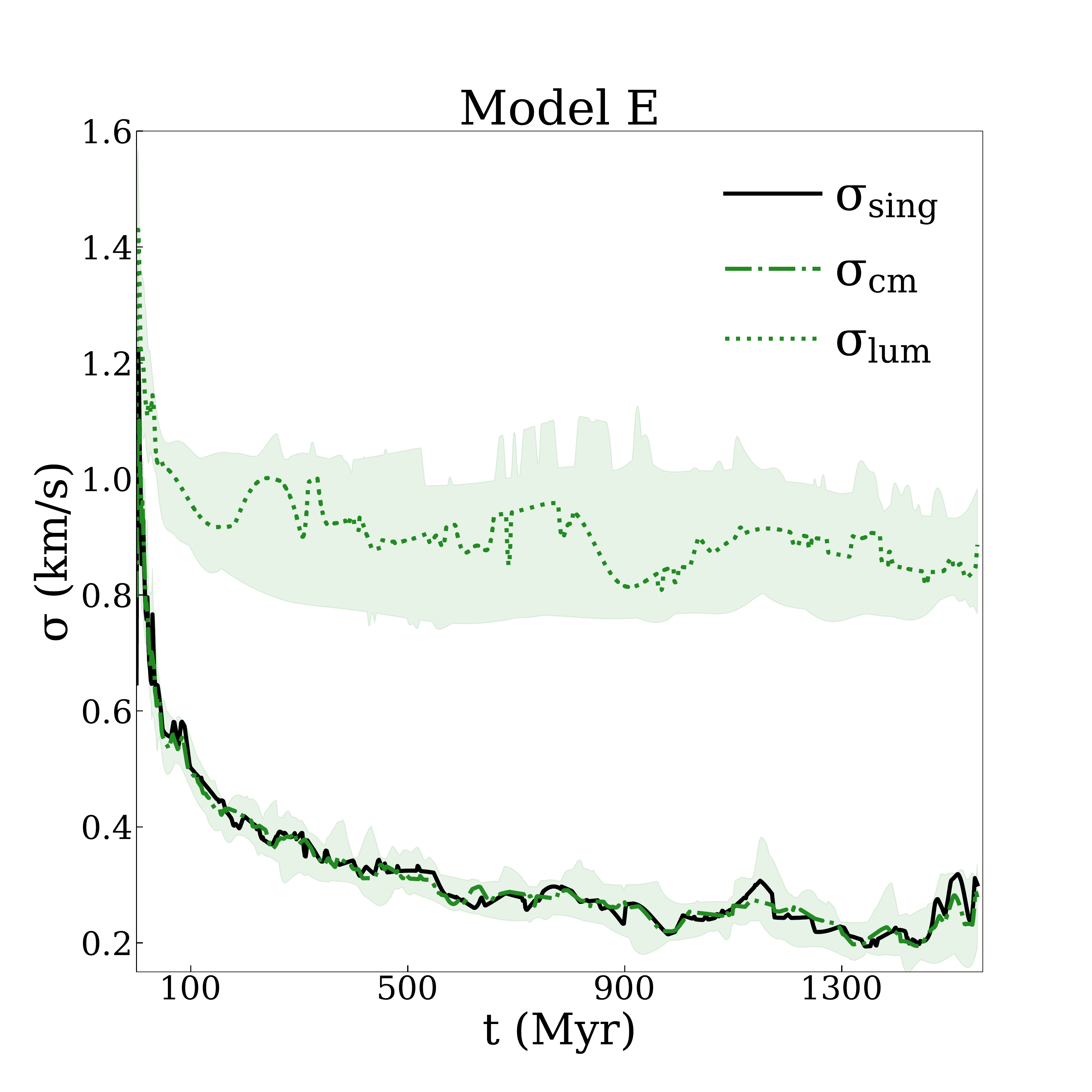}
\caption{Zoom of Fig. \ref{fig:allsigmaev} for model B and E showing
the evolution of $\sigma_{sing}$ (black line), $\sigma_{cm}$ (dot-dashed line) and
$\sigma_{lum}$ (dotted line).
The behaviours of $\sigma_{cm}$ and $\sigma_{sing}$ overlap each other in both the models.}
\label{fig:zoom}
\end{figure}

For the sake of clarity Figure \ref{fig:zoom} displays a comparison among
$\sigma_{sing}$, $\sigma_{lum}$ and $\sigma_{cm}$  for model B (left panel) and
E (right panel) that shows the overlap between $\sigma_{cm}$ and
$\sigma_{sing}$. This outcome is typical of any model as anticipated by the
results of Sect. \ref{sigma}. Cluster models on an initial virial equilibrium
(A, B and C) show lower values (a factor 2) of $\sigma$ with respect to
initially sub-virial models (D, E and F).

We notice a general decreasing of the velocity dispersion with time,
independently on the method, that is common to all the models studied. The
velocity dispersion decreases as consequence of the evolution of the clusters
that, as discussed in section \ref{ev}, undergo mass loss due to stellar
evolution and dynamics in form of escaping stars.

\subsection{Dynamical Mass Estimates}
\label{mass}

The simplest way to estimate the mass of a star cluster is by mean of the
assumption of virial equilibrium, $Q=1$. In this case the \textit{dynamical}
(or \textit{virial}) mass $M_{d}$ of a star cluster can be estimated with the
following relation \citep{spi87}:

\begin{equation}
\label{eq:5}
M_{d}=\frac{ \eta\, \sigma_{1D}^{2} \, R_{eff}}{G}
\end{equation}

where $\sigma_{1D}$ is the (1D, i.e. along the line of sight) velocity
dispersion, $G$ is the gravitational constant, $R_{eff}$ is the \vir{effective}
radius of the cluster, usually
assumed as the half light radius of the systems \citep{gieles2010}, and $\eta $
is a dimensionless  factor which depends on the cluster density profile (in the
ideal case of a self gravitating homogeneous sphere of radius $R_{eff}$ in
energy equipartition, $\eta =5$).

\noindent  It is relevant noting that the assumption of virial
equilibrium is  questionable for open star clusters in our Galaxy  because of
the tidal galactic field which can be relevant on the sizes of these star
clusters (through differential rotational velocity).

Actually, what likely affects most the dynamical mass estimate is the presence
of binary stars in the stellar system, because they would naturally induce a
bias of $\sigma$  due to their pair orbits may increase significantly the value
of $\sigma$ which, once inserted in Eq. \ref{eq:5}, leads to an overestimate
even if the assumption of global virial equilibrium is valid.

Using our simulations of isolated open clusters containing an evolving binary
population we can make a straightforward comparison between the different
dynamical mass estimates derived by each of the methods described before.  The
comparison is done by measuring the various ratios between the $\sigma^{2}$
derived with each method and the $\sigma^{2}_{cm}$ which, in its turn, would be
the one giving the correct evaluation of mass.

The results of this analysis are summarized in Table \ref{sigma_frac_tot}
for the three luminosity-limited samples.

\begin{table}[]
\begin{tabular}{cccccccccc}
\hline
\multicolumn{1}{l}{}                        & \multicolumn{3}{c}{\multirow{2}{*}{\textit{Sample2}}}                                      & \multicolumn{3}{c}{\multirow{2}{*}{\textit{Sample1}}}                                      & \multicolumn{3}{c}{\multirow{2}{*}{\textit{Sample0}}}                 \\
\multicolumn{1}{l}{}                        & \multicolumn{3}{c}{}                                                              & \multicolumn{3}{c}{}                                                              & \multicolumn{3}{c}{}                                         \\ \hline
\multicolumn{1}{c|}{\multirow{2}{*}{Model}} & \multirow{2}{*}{$\sigma_{tot}^{2}/\sigma_{cm}^{2}$} & \multirow{2}{*}{ $\sigma_{lum}^{2}/\sigma_{cm}^{2}$} & \multicolumn{1}{c|}{\multirow{2}{*}{$\sigma_{sing}^{2}/\sigma_{cm}^{2}$}} & \multirow{2}{*}{$\sigma_{tot}^{2}/\sigma_{cm}^{2}$} & \multirow{2}{*}{ $\sigma_{lum}^{2}/\sigma_{cm}^{2}$} & \multicolumn{1}{c|}{\multirow{2}{*}{$\sigma_{sing}^{2}/\sigma_{cm}^{2}$}} & \multirow{2}{*}{$\sigma_{tot}^{2}/\sigma_{cm}^{2}$} & \multirow{2}{*}{ $\sigma_{lum}^{2}/\sigma_{cm}^{2}$} & \multirow{2}{*}{$\sigma_{sing}^{2}/\sigma_{cm}^{2}$} \\
\multicolumn{1}{c|}{}                       &                    &                    & \multicolumn{1}{c|}{}                   &                    &                    & \multicolumn{1}{c|}{}                   &                    &                    &                    \\ \hline
\multicolumn{1}{c|}{}  &        &        & \multicolumn{1}{c|}{}      &        &        & \multicolumn{1}{c|}{}      &         &          &        \\
\multicolumn{1}{c|}{A} & 16.2 & 2.5  & \multicolumn{1}{c|}{1.020}  & 36.5 & 4.4   & \multicolumn{1}{c|}{1.00}  & 38.6 & 3.9 & 0.9  \\
\multicolumn{1}{c|}{}  &        &        & \multicolumn{1}{c|}{}      &        &        & \multicolumn{1}{c|}{}      &         &          &        \\
\multicolumn{1}{c|}{B} & 75.2 & 7.3 & \multicolumn{1}{c|}{1.039} & 157.8 & 17.4 & \multicolumn{1}{c|}{0.918} & 151.1 & 21.7 & 0.8  \\
\multicolumn{1}{c|}{}  &        &        & \multicolumn{1}{c|}{}      &        &        & \multicolumn{1}{c|}{}      &         &          &        \\
\multicolumn{1}{c|}{C} & 269.6 & 23.4 & \multicolumn{1}{c|}{1.195} & 533.5 & 52.1 & \multicolumn{1}{c|}{1.029} & 687.2 & 52.1 & 0.8  \\
\multicolumn{1}{l|}{}                       & \multicolumn{1}{l}{} & \multicolumn{1}{l}{} & \multicolumn{1}{l|}{}                   & \multicolumn{1}{l}{} & \multicolumn{1}{l}{} & \multicolumn{1}{l|}{}                   & \multicolumn{1}{l}{} & \multicolumn{1}{l}{} & \multicolumn{1}{l}{} \\ \hline
\multicolumn{1}{c|}{}  &        &        & \multicolumn{1}{c|}{}      &        &        & \multicolumn{1}{c|}{}      &         &          &        \\
\multicolumn{1}{c|}{D} & 38.6 & 4.5 & \multicolumn{1}{c|}{1.000}  & 42.5 & 6.5  & \multicolumn{1}{c|}{1.000}  & 45.7 & 6.6 & 1.0  \\
\multicolumn{1}{c|}{}  &        &        & \multicolumn{1}{c|}{}      &        &        & \multicolumn{1}{c|}{}      &         &          &        \\
\multicolumn{1}{c|}{E} & 115.2 & 10.2 & \multicolumn{1}{c|}{ 1.239} & 295.4 & 31.6 & \multicolumn{1}{c|}{0.953} & 245.3 & 21.8 & 0.9  \\
\multicolumn{1}{c|}{}  &        &        & \multicolumn{1}{c|}{}      &        &        & \multicolumn{1}{c|}{}      &         &          &        \\
\multicolumn{1}{c|}{F} & 504.1 & 44.7 & \multicolumn{1}{c|}{0.953} & 841.0 & 82.7 & \multicolumn{1}{c|}{0.781} & 1060.8 & 84.2 & 0.5  \\
\multicolumn{1}{c|}{}  &        &        & \multicolumn{1}{c|}{}      &        &        & \multicolumn{1}{c|}{}      &         &          &        \\ \hline
\end{tabular}
\caption{Ratios of various determinations of $\sigma^2$ respect to $\sigma_{cm}^2$, which would be the correct one to provide the cluster dynamical mass, in \textit{sample2}, \textit{sample1} and \textit{sample0}.
}
\label{sigma_frac_tot}
\end{table}

The overestimate of the dynamical mass produced by method 1 is large in all our
models and samples.  Actually, we find
$16.2<\sigma_{tot}^{2}/\sigma_{cm}^{2}<504.1 $ for \textit{sample2}, $
36.5<\sigma_{tot}^{2}/\sigma_{cm}^{2}<841 $ for \textit{sample1} and $
38.6<\sigma_{tot}^{2}/\sigma_{cm}^{2}<1060.8$ for \textit{sample0}.  As expected, the
overestimate increases with the fraction of binaries. Moreover, the bias is
greater for the sub-virial models D, E and F. This difference reflects their
larger fraction of binaries as estimated at $1.5$ Gyr (see Tab.
\ref{frac_tot}). Since models on an initial sub$-$virial state show a high
percentage of binaries, their velocity dispersion $\sigma_{tot}$ is larger
yielding to a significant overestimate of $M_{d}$ with respect to models with
a lower fraction of binaries.

A similar outcome is found when considering $\sigma^{2}_{lum}$. The
overestimate of the mass reflects the corresponding velocity dispersion variation, $2.5<\sigma_{lum}^{2}/\sigma_{cm}^{2}<44.7 $ for \textit{sample2}, $
4.4<\sigma_{lum}^{2}/\sigma_{cm}^{2}<82.7 $ for \textit{sample1} and $
3.9<\sigma_{lum}^{2}/\sigma_{cm}^{2}<84.2 $ for \textit{sample0}.  As already
mentioned, method 3 is the method that better mimics what observations give for
measures of the velocity dispersion of a star cluster. Thus, the results reported
in column 2 of each box of Table \ref{sigma_frac_tot} provide a reliable
estimate of the correction to apply to observations to account for the binary
population of the stellar system.

On the other side when we estimate the ratio
$\sigma^{2}_{sing}/\sigma^{2}_{cm}$ we obtain values very close to 1. This
outcome reflects what we derived in the previous Sect. \ref{sigma}.

In Table \ref{fit} we report, for each sample, the parameters of log-linear
fits of the results of Tab. \ref{sigma_frac_tot}:

\begin{equation}
\label{eq:6}
\log\, (\sigma_{i}/\sigma_{cm})^{2}    = a_i  \log f_{b}  +  b_i,
\end{equation}

where $\sigma_{i}$ for $i=1,2,3$ corresponds, respectively, to $\sigma_{tot}$, $\sigma_{lum}$
and $\sigma_{sing}$.

Now, recalling Eq. \ref{eq:5}, we can see that the value of the dynamical mass
as estimated by observations is actually

\begin{equation}
\label{eq:7}
\log\dfrac{M_{d,obs}}{M_{d}}=\log\left(\dfrac{\sigma_i}{\sigma_{cm}}\right)^2,
\end{equation}

where $\log (\sigma_{i}/\sigma_{cm})^{2}$ come from Eq. \ref{eq:6}.  The
formula above is a good correction formula to use to have a proper estimate of
the virial mass from observed values of velocity dispersion biased by the
presence of a binary population.

Given the values in Table \ref{sigma_frac_tot}, and taking as most representative
sample the deepest in luminosity (\textit{sample2}), and as most likely $\sigma$
determination that we call $\sigma_{lum}$, we have that the logarithmic ratio
between the observational and correct dynamical mass estimate varies in the
range $0.4 \leq M_{d,obs}/M_{d} \leq 1.4$ for $0.08\leq f_b \leq 0.40$ in the
virial models (A, B, C) and in the range $0.7\leq  M_{d,obs}/{M_{d}}  \leq 1.7$
for $0.09 \leq f_b \leq 0.42$ in the sub-virial models (D, E, F).
This means an overstimate from a factor 2.5 to a factor 45, which is absolutely non-negligible.

\begin{table}[]
\begin{tabular}{@{}ccccc|cccc|cccc|cccc|cccccccl@{}}
\multicolumn{1}{l}{}                                     & \multicolumn{8}{c|}{\textit{Sample2}}                                                    & \multicolumn{8}{c|}{\textit{Sample1}}                                                             & \multicolumn{8}{c}{\textit{Sample0}}                                                                                               \\ \cmidrule(l){2-25}
                                                         & \multicolumn{4}{c|}{Model A, B, C}     & \multicolumn{4}{c|}{Model D, E, F}              & \multicolumn{4}{c|}{Model A, B, C}              & \multicolumn{4}{c|}{Model D, E, F}              & \multicolumn{4}{c|}{Model A, B, C}                                   & \multicolumn{4}{c|}{Model D, E, F}                          \\ \cmidrule(l){2-25}
                                                         &  & \textit{a} & \textit{b} & \textit{} & \textit{} & \textit{a} & \textit{b} & \textit{} & \textit{} & \textit{a} & \textit{b} & \textit{} & \textit{} & \textit{a} & \textit{b} & \textit{} & \textit{} & \textit{a} & \textit{b} & \multicolumn{1}{c|}{\textit{}} & \textit{} & \textit{a} & \textit{b} & \multicolumn{1}{l|}{} \\ \midrule
\multicolumn{1}{c|}{}                                    &  &            &            &           &           &            &            &           &           &            &            &           &           &            &            &           &           &            &            & \multicolumn{1}{c|}{}          &           &            &            & \multicolumn{1}{l|}{} \\
\multicolumn{1}{c|}{$\sigma^{2}_{tot}/\sigma^{2}_{cm}$}  &  & 1.71       & -0.95      &           &           & 1.52       & 0.18       &           &           & 1.8        & -0.53      &           &           & 1.8        & -0.3       &           &           & 2.34       & -2.18      & \multicolumn{1}{c|}{}          &           & 2.17       & -1.22      & \multicolumn{1}{l|}{} \\
\multicolumn{1}{c|}{}                                    &  &            &            &           &           &            &            &           &           &            &            &           &           &            &            &           &           &            &            & \multicolumn{1}{c|}{}          &           &            &            & \multicolumn{1}{l|}{} \\
\multicolumn{1}{c|}{$\sigma^{2}_{lum}/\sigma^{2}_{cm}$}  &  & 1.51       & -2.51      &           &           & 1.53       & -2.18      &           &           & 2          & -3.5       &           &           & 1.9        & -2.9       &           &           & 1.68       & -2.29      & \multicolumn{1}{c|}{}          &           & 1.83       & -2.48      & \multicolumn{1}{l|}{} \\
\multicolumn{1}{c|}{}                                    &  &            &            &           &           &            &            &           &           &            &            &           &           &            &            &           &           &            &            & \multicolumn{1}{c|}{}          &           &            &            & \multicolumn{1}{l|}{} \\
\multicolumn{1}{c|}{$\sigma^{2}_{sing}/\sigma^{2}_{cm}$} &  & 0.03       & -0.06      &           &           & -0.02      & 0.07       &           &           & 0.05       & -0.17      &           &           & 0.01       & -0.04      &           &           & -0.03      & -0.04      & \multicolumn{1}{c|}{}          &           & -0.38      & 0.9        & \multicolumn{1}{l|}{} \\
\multicolumn{1}{c|}{}                                    &  &            &            &           &           &            &            &           &           &            &            &           &           &            &            &           &           &            &            & \multicolumn{1}{c|}{}          &           &            &            & \multicolumn{1}{l|}{} \\ \bottomrule
\end{tabular}
\caption{Values of the parameters in the log-linear fitting formula Eq. \ref{eq:6} of the results of Table \ref{sigma_frac_tot} for each sample and each model.
}
\label{fit}
\end{table}

\section{Summary and Conclusions}
In this work we have presented a large suite
of numerical simulations performed with the high precision, direct summation, \texttt{NBODY7} code with the aim to investigate the effect of the presence of binary stars in the determination of the dynamical mass of stellar systems.

In particular, we focused our attention to models of
Galactic open clusters, since these systems harbour abundant populations  of binary stars, and are made of a relatively small number of stars, which makes numerical simulations affordable in terms of computational effort and, hence, allows an easier exploration of the parameter space.

In this study, we considered clusters containing, initially, $1000$ stars, spanning a wide range of initial conditions, including different
primordial binary fractions ($5$\%, $15$\% and $30$\%) and initial virial ratio ($Q_0=2K_0/|W_0|$) $Q_0=0.5$ and $Q_0=1$. We followed the evolution of each model up to $1.5$ Gyr. Our simulations neglected the effect of the tidal field of the Milky Way, which we plan to include in the future.

The time evolution of the various models mass and half mass radius
were as expected: the mass decreases in all models, while the half mass
radius increases because of the combined effects of stellar evolution and
two/three body encounters that produce escapers.  As expected, in sub-virial ($Q_0=0.5$) models the mass loss is more significant than in initially virialised ($Q_0=1$) systems.

In addition, for each model we looked at the internal velocity dispersion. In detail, we normalized each estimate of the velocity dispersion ($\sigma_{sing}$, $\sigma_{tot}$, and
$\sigma_{lum}$) to $\sigma_{cm}$, this latter being the one which best represents the actual kinetic content of the cluster, so it would be the proper one to evaluate a virial mass. The various estimates of the velocity dispersion
we used have the aim to reproduce what observers obtain as estimates of the velocity dispersion of a star cluster.

Independently of the adopted initial model and of the specific velocity
dispersion estimate considered, a clear trend emerges of larger velocity
dispersion at larger binary fractions.  This, in turn, produces an
overestimate of the cluster dynamical mass when computed using blindly
Eq. \ref{eq:5}. The overestimate depends on the way the velocity dispersion is derived. For reasonable values of the actual binary percentage (8\% to 42\%) it can be up to a factor of $45$. This implies that neglecting in part or completely the binary population in a cluster has profound impact in the total mass estimate.

To take the binary effect into account, we provide in Sect. \ref{mass} fitting formulae which can be used to correct the cluster mass evaluation whenever some estimate of the binary fraction is available.

This has an impact on Galactic open clusters which is anyway limited by the increasing precision of observational data which, nowadays, makes it possible to infer the binary fraction and the mass with enough precision from photometry only \citep{Sele2017,Boro2019}. However, when considering other stellar systems, like dwarf galaxies in the Local Group, it is clear that a quantitative insight of the overestimate of the velocity dispersion caused by the binary population together with the assumption of virialisation could be extremely helpful to determine the quantity of dark matter present. We are aware that the application of the present results to dwarf spheroidal galaxies can be done just in a tentative way, because the primordial binary fraction and their evolution due to the internal dynamics are, likely, significantly different from those in open clusters.

\bibliographystyle{aasjournal}

\bibliography{rasetalBIN}

\end{document}